\documentclass{jfm}
\usepackage{graphicx}
\usepackage{epstopdf, epsfig}
\usepackage{bm}
\usepackage{amsmath} 
\usepackage{subcaption}
\usepackage[figuresright]{rotating}
\usepackage{diagbox}
\usepackage{amssymb}
\usepackage{siunitx}
\usepackage{soul}
\usepackage{xcolor}

\shorttitle{A Coupled Model of Liquid Droplet Impact on Solid Material Surfaces}
\shortauthor{Hao et al.}

\title{An Analytical-Numerical Coupled Model of Liquid Droplet Impact on Solid Material Surfaces}

\author{Hao Hao\aff{1}\corresp{\email{hao.hao17@imperial.ac.uk}}, Maria N. Charalambides\aff{1}, Yannis Hardalupas \aff{1}, Antonis Sergis\aff{1} and Alex M. K. P. Taylor\aff{1} 
}

\affiliation{\aff{1}Department of Mechanical Engineering, Imperial College London, London SW7 2AZ, United Kingdom}

\begin{document}

\maketitle

\begin{abstract}
Impacts of liquid droplets on wind turbine blade surfaces, for example sea sprays, can result in material damage through erosion. In this study, we derive an explicit, closed-form analytical approximation for droplet impact and subsequent spreading on a solid surface in inertia-dominated regimes of large Reynolds and Weber numbers. The formulation extends an existing theoretical framework based on inviscid potential flow for a rising expanding disk in an infinite liquid domain. The modified solution provides full spatio-temporal pressure distributions and impact force histories on the impact surface over the entire impact duration, capturing both the early-time self-similar flow and the inertia-driven lamella spreading following the peak impact force. The predicted pressure and force profiles show good agreement with analytical, numerical and experimental results reported in the literature, including accurate reproduction of the well-known ring-shaped pressure distribution. Key quantities\textemdash such as the radial location and magnitude of peak pressure, as well as the timing and magnitude of the peak impact force\textemdash are predicted analytically with reasonable accuracy. To enable solid material erosion analysis, the analytical liquid-phase solution is coupled with a finite-element (FE) simulation for the solid response. This analytical-numerical coupled method (ANCM) eliminates the need to explicitly simulate droplet fluid dynamics, which is conventionally performed using smoothed particle hydrodynamics (SPH). As a result, for the purpose of material response analysis, the proposed approach achieves grid independence at substantially lower mesh resolutions and reduces computational cost by more than \qty{97}{\percent} compared to SPH-based simulations, while maintaining or improving numerical accuracy.
\end{abstract}

\begin{keywords}
Authors should not enter keywords on the manuscript, as these must be chosen by the author during the online submission process and will then be added during the typesetting process (see http://journals.cambridge.org/data/\linebreak[3]relatedlink/jfm-\linebreak[3]keywords.pdf for the full list)
\end{keywords}

\section{Introduction}\label{1}
Impacts between liquid droplets and solid surfaces can commonly cause erosion problems in engineering applications, including aircraft surface erosion \citep{Gohardani1}, steam turbine blade erosion \citep{Preece1} and wind turbine blade leading-edge erosion \citep{Nick1}. More recently on wind turbines, impact between droplets, in particular sea sprays or rain droplets, and blade surfaces of wind turbines can damage blades due to leading-edge erosion (LEE), and this has become an increasingly serious problem on longer blades and offshore turbines due to the higher-speed impacts at the blade tip \citep{Verma1}. Impacts lead to material removal from the blade, aerofoil geometry damage and even failure in the absence of appropriate maintenance \citep{Yang1}, which considerably reduces energy production rates and eventually the lifetime of turbine blades. For these reasons, coating protections against liquid-drop impact erosion have become the highest priority for all blade manufacturers \citep{Herring1} and have motivated research on liquid-solid impact studies for decades.

In the pioneering experimental work on single droplet impact to solid materials, \cite{Field3} studied the relation between stress wave propagation in the solids, which react and reinforce to generate material damage, and the pressure loadings of liquid impact. In brittle materials, they found the main characteristics of damage on the impact surface is a ring fracture surrounding a relatively less damaged or even undamaged area. The same ring pattern had also been observed in previous experiments by \cite{Engel1} and, almost at the same time, by \cite{Brunton1}. \cite{Brunton1} explained the ring erosion pattern due to the non-uniformity of surface loadings where high tensile forces, due to the lateral motion of spreading lamella \citep{Taylor1}, developed at the edge of the loaded area during impact \citep{Field3} while the central region of the surface was elastically compressed.

In more recent decades, numerical fluid-structure interaction (FSI) models have been developed to explore liquid-solid impacts. In the context of LEE on wind turbine blades, \cite{Keegan1} has employed and compared two dominant finite-element (FE) approaches: coupled Eulerian-Lagrangian (CEL) and smoothed particle hydrodynamics (SPH). Both approaches are based on the FE method which is advantageous regarding the solid material analysis \citep{Doagou1}, however, suffers from large computational demands due to large deformation and small-scale interface evolutions of the liquid phase, or alternatively, poor resolutions on the liquid phase \citep{Zhou2}. It is found that CEL is more computationally expensive than SPH in simulating droplet impact problems \citep{Doagou1}. Alternatively, there are coupled methods that involve a separate computational fluid dynamics (CFD) simulation and a numerical FE analysis for the liquid and solid phases respectively. \cite{Amirzadeh1, Amirzadeh2} used an in-house CFD model to calculate the pressure loadings upon a single droplet impact assuming rigid solid surface, and exerted the obtained pressure profiles onto the surface of the solid materials in an FE analysis. The coupled method benefits from better accuracy of the droplet dynamics, but also requires high computational power and, as a new consideration, a deliberate design of grids at the interface for data exchange. The latter is not an issue though if the two phases can be meshed in the same solver, such as in the recent work by \cite{Nick1} who coupled a level-set method and an FE analysis for the multi-phase problem in Comsol Multiphysics\raisebox{1ex}{\scriptsize \textregistered}. 

Analytical research on liquid-solid impact traces back to $1920$s investigating impact loadings on the contact surface \citep{Cook1, Karman1}. Throughout this paper, variables are made dimensionless using $R_0$, $U_0$ and $\rho_l$, with $R_0$ the droplet initial radius\footnote{In the case of wedge-shaped object, droplet radius is not applicable and an arbitrary length-scale should be used instead.}m $U_0$ the impact speed and $\rho_l$ the liquid density; and lower-case letters denote dimensionless forms of corresponding upper-case variables, unless specified otherwise. In the pioneering work, \cite{Wagner1} first proposed mathematical models of the liquid-solid impact problem. The two-dimensional mathematical models introduced the concept of \textit{inner} and \textit{outer} regions for the impact pressure distributions regarding the wet radius $a(t)$ that expands in time, $t$. For radius less than $a(t)$ (named `outer region' \footnote{The names refer to the relative positions to the wet radius; see \cite{Wagner1} for details.}), the pressure profile, $p_{outer}(x,t)$, refers to the impact pressure under the droplet and comes from the unsteady Bernoulli pressure to the leading order:
\begin{equation}
  p_{outer}(x,t)=\frac{1}{\sqrt{1-\frac{x^2}{a^2}}}\frac{da}{dt},
\end{equation}
at dimensionless distance $x$ from the impact centre on the surface, while at radius beyond $a(t)$ (named `inner region'), the profile, $p_{inner}(x,t)$, refers to pressure in the spreading lamella and is in the parametric form:
\begin{equation}
  p_{inner}(\tau)=\frac{2\sqrt{-\tau}}{(1+\sqrt{-\tau})^2},
\end{equation}
with parameter $\tau$ implicitly defined from a conformal mapping of $x$ and $a(t)$. Since then, solutions have been further developed in other studies \citep{Tassin1}, including inclusion of the steady Bernoulli pressure \citep{Logvinovich1}, extension to axi-symmetric solutions \citep{Scolan1} and modification to droplet impact onto flat solid surface \citep{Oliver1, Negus1}. However, impact flow solutions from Wagner's theoretical droplet impact framework are mainly limited to early impact time.

From the literature, we can broadly divide the droplet impact into four main regimes:
\begin{itemize}
    \item Pressure-driven self-similar impact: initial impact develops a self-similar region on the impact surface \citep{Philippi1}. In this regime, flow outside of the self-similar region continues falling with the original impact velocity until the high pressure region has spread over the whole droplet at time around $T=R_0/U_0$ (where $T$, $R_0$ and $U_0$ denote the dimensional time, droplet radius and impact speed, respectively)\citep{Egger1,Gordillo2,Lagubeau}. Approaching the end of this regime, the original falling flow is hence effectively reduced, and we observe that pressure within the flow drops rapidly \citep{Egger1}, and the impact force deviates from the self-similar theory \citep{Gordillo2,Hao1};
    \item Inertia-driven self-similar spreading: the spreading of drop is driven by drop inertia in the radial direction \citep{Cheng}. As found by \cite{Egger1} and others \citep{Yarin,Roisman1,Roisman2}, the flow evolution can be nicely approximated by a time-dependent `hyperbolic' flow. During this regime, the impact force on the solid surface decays from the peak value to almost zero \citep{Gordillo2,Cheng,Mitchell};
    \item Maximum spreading (before breakage): the droplet eventually reaches its maximum spreading diameter, limited by surface tension and viscosity \citep{Lagubeau,Sanjay,Cheng}. As the axial momentum of impacting drops approaches zero near the maximum spreading diameter, the impact force will be very weak \citep{Cheng};
    \item Splashing, pinning, retraction and jump-off: depending on the Weber number $We$ and the Ohnesorge number $Oh$, further physical phenomena may happen \citep{Egger1,Sanjay,Garcia2020}, particularly for hydrophobic surfaces \citep{Sanjay2,Sanjay3}. However, in the current study, we limit our focus to non-hydrophobic surfaces, where impact pressure on the wall and corresponding force accompanying these phenomena are too small to be of any concern (to impact erosion).
\end{itemize}
From the descriptions, the first two regimes are particularly important to impact erosion for their key contribution to impact loadings on the solid surface and the fact that the maximum impact force occurs during the transition from the first to the second regime \citep{Hao1}. However, as noted by \cite{Gordillo2}, there is no existing theoretical understanding that could bridge the first two self-similar regimes. One of the most promising attempts was given by \cite{Mitchell} who empirically interpolated the two regimes using an exponential function. Nevertheless, the obtained empirical solution only provides the temporal evolution of force profile without a spatio-temporal pressure distribution. More recently, the present authors introduced the \textit{separation point} of the flow to Wagner's theoretical droplet impact framework \citep{Hao1}, which was shown as a better prediction to the wet radius.
The concept of separation point helped extend the analytical solution up to intermediate times (as the end of the first impact regime) and left a gap for later times and fluid-solid coupling, which this work now addresses.

From the above literature review, we see the need to develop an explicit analytical solution in closed form that can reasonably describe the droplet impact and spreading phases, which are of primary interest in engineering applications. Besides, there remains a notable gap in the literature on liquid impact erosion studies that couple analytical droplet-impact models with numerical FE analyses of the solid response. Such a coupling has the potential to substantially reduce the high computational cost typically associated with fully numerical approaches, while maintaining, or even improving, the accuracy of the simulation results. Accordingly, the present study aims to address these gaps. Section \ref{2} introduces the theoretical framework for liquid droplet impact and derives an analytical solution for the impact pressure. Section \ref{3} presents a novel coupled modelling approach that integrates the analytical liquid solution with numerical simulations of the solid material. In section \ref{4}, the coupled model is validated through comparisons with the conventional SPH method. Section \ref{5} discusses the accuracy and computational advantages of the proposed approach, and section \ref{6} concludes the study.

\section{Theories on Liquid Droplet Impact}\label{2}
In this section, we extend the early-to-intermediate-time solution from \cite{Hao1} beyond $t\approx0.27$ to a full-duration analytic approximation by implementing new assumptions. To clarify the contributions of the present study relative to the previous study, we note that section \ref{2.1} recapitulates the governing equations and early-to-intermediate-time solution from \cite{Hao1}; section \ref{2.2} presents new evaluations and two simplifying assumptions\textemdash where any unavoidable reuse of results from the prior study is explicitly cited; while section \ref{2.3} and subsequent sections introduce the new (modified) solution and its advancements, including the liquid-solid coupling in section \ref{3}.

\subsection{Mathematical Model}\label{2.1}
\subsubsection{Governing equations}\label{2.1.1}
\begin{figure}
    \centering
    \begin{subfigure}[b]{0.48\textwidth}
        \caption{}
        \includegraphics[width=\textwidth]{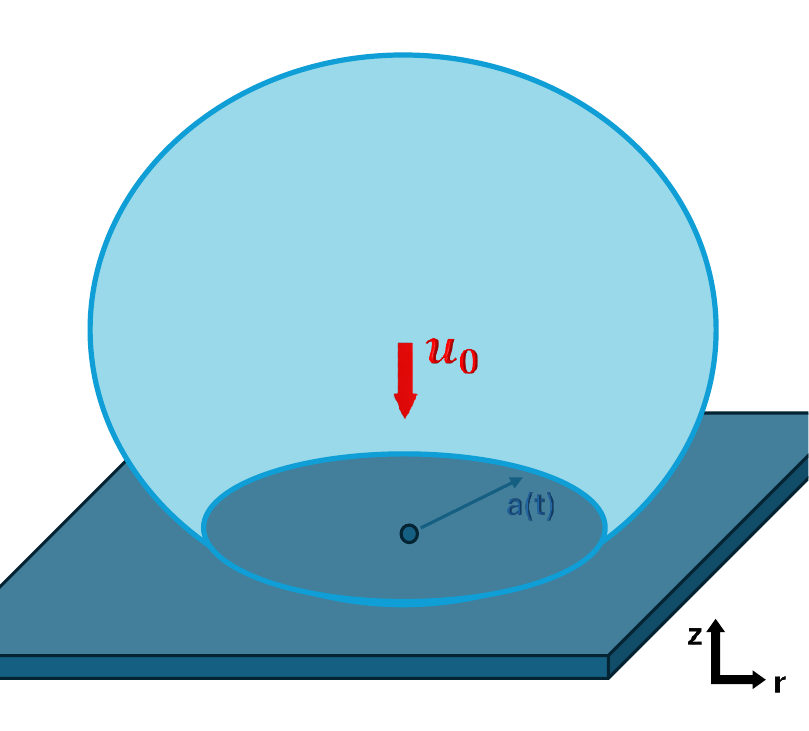}
        \label{figure_disk&drop1}
    \end{subfigure}
    \hfill
    \begin{subfigure}[b]{0.48\textwidth}
        \caption{}
        \includegraphics[width=\textwidth]{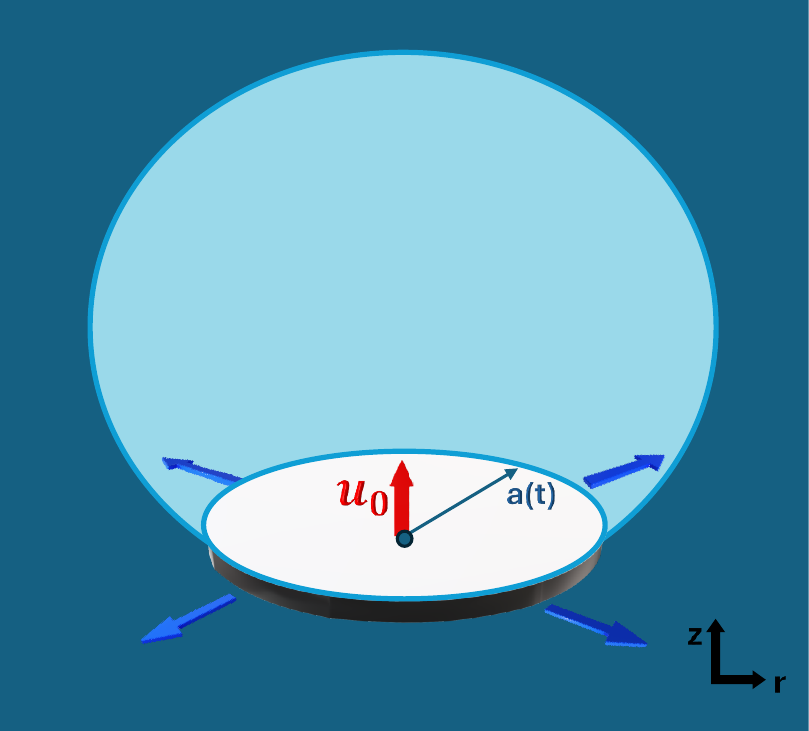}
        \label{figure_disk&drop2}
    \end{subfigure}
    \caption{The problem of liquid droplet impact (downwards at speed $u_0$) on a solid surface (a) has been found to be analogous to the problem of a thin rising expanding solid circular disk (upwards at speed $u_0$) in an infinite mass of liquid (b). The blue arrows in (b) denote the radial expansion of the disk with the wet radius. The figure is reproduced after \cite{Philippi1} and \cite{Hao1}.}
\label{figure_disk&drop}
\end{figure}

\cite{Hao1} has introduced the mathematical framework of a flat rising expanding disk in infinite water. Flows within such framework are found to be analogous to liquid droplet impact on flat solid surface (see figure \ref{figure_disk&drop}). The mathematical problem describes the motion of a thin circular disk of radius $a(t)$ with constant velocity $\bm{u_0}$ normal to its plane in an infinite mass of liquid. It is worth-noting that the expanding disk radius $a(t)$ in time introduces unsteadiness to the mathematical problem, which is the key difference to the original Lamb's steady disk framework \citep{Lamb1}. For the described problem, the flow is an incompressible and inviscid axisymmetric ideal fluid field with far field at rest (hence irrotational), so the time-dependent velocity potential $\phi(r,z,t)$ can be defined:
\begin{equation}
  \bm{u}(r,z,t)=\nabla\phi(r,z,t),
\end{equation}
where time is defined at the initial contact, and, in the inertia moving frame of reference at velocity $\bm{u_0}$, the origin is taken at the centre of the disk. By the mass conservation law, $\phi(r,z)$ satisfies the Laplacian equation, written in cylindrical coordinates: 
\begin{equation}
  \nabla^2\phi=\frac{1}{r} \frac{\partial}{\partial r} \left( r \frac{\partial\phi}{\partial r} \right) + \frac{\partial^2 \phi}{\partial z^2}=0,
  \label{Lap}
\end{equation}
and the Bernoulli's conservation equation everywhere in the liquid:
\begin{equation}
  \frac{\partial \phi}{\partial t} + \frac{1}{2} |\nabla \phi|^2 + \frac{p}{\rho} = b(t),
  \label{Ber}
\end{equation}
where the dimensionless density $\rho$ is $1$ and the Bernoulli `constant' $b(t)$ is a function of time only. The ideal fluid field in the control volume satisfies three boundary conditions, namely:
\begin{equation}
    \frac{\partial \phi}{\partial z}(r,0,t)=1  \;\;\;\;\;\;\;\; \text{for $r \leq a(t)$, $t \geq 0$},
    \label{BC1}
\end{equation}
due to non-permeability at the disk surface $z=0$; and:
\begin{equation}
    \phi(r,0,t)=0 \;\;\;\;\;\;\;\; \text{for $r > a(t)$, $t \geq 0$},
    \label{BC2}
\end{equation}
and:
\begin{equation}
    \nabla \phi(r,z,t)\rightarrow0  \;\;\;\;\;\;\;\; \text{as $r^2+z^2 \rightarrow \infty$, $t \geq 0$},
    \label{BC3}
\end{equation}
since velocity field is undisturbed far from the disk surface.

\subsubsection{Solutions on the solid surface}\label{2.1.2}
\cite{Hao1} has analytically derived the solutions of the flow, described by equations \ref{Lap} and \ref{Ber} subject to boundary conditions \ref{BC1}, \ref{BC2} and \ref{BC3}, on the solid disk surface ($z=0$). Particularly, solutions on the solid surface are all we need for the interest of solid material analysis upon liquid droplet impact. The solutions include the velocities:
\begin{equation}
  \bm{u}(r,0,t) = \nabla \phi(r,0,t) = (\frac{2r}{\pi\sqrt{a(t)^2-r^2}}) \bm{\underline{\hat{r}}} + \bm{\underline{\hat{z}}} \;\;\;\;\;\;\;\; \text{for $r < r_b(t) <a(t)$} ,
\label{sol-u}
\end{equation}
for $\bm{\underline{\hat{r}}}$ and $\bm{\underline{\hat{z}}}$ as the unit vectors in $r$ and $z$ directions, and the unsteady flow potential:
\begin{equation}
  \frac{\partial \phi}{\partial t}(r,0,t) =- \frac{2}{\pi} \frac{a(t)a'(t)}{\sqrt{a(t)^2-r^2}} \;\;\;\;\;\;\;\; \text{for $r < r_b(t) < a(t)$},
\label{sol-dphidt}
\end{equation}
where $r_b$ is the radial boundary of the solutions, beyond which the disk-analogy breaks down.
  
By equation \ref{Ber} and solutions of equations \ref{sol-u} and \ref{sol-dphidt}, the Bernoulli pressure on the solid surface upon droplet impact are further derived:
\begin{equation}
  p(r,0,t) = b(t) - \frac{2}{\pi^2} \frac{r^2}{a(t)^2-r^2} + \frac{2}{\pi} \frac{a(t)a'(t)}{\sqrt{a(t)^2-r^2}} \;\;\;\;\;\;\;\; \text{for  $r <  r_b(t) < a(t)$},
\label{sol-p}
\end{equation}
where the Bernoulli constant has been commonly neglected in the literature \citep{Philippi1, Tassin1} due to the relatively lower order of magnitude. By integrating the impact pressure of equation \ref{sol-p} over the solid surface, impact force $f(t)$ upon droplet impact can be obtained, in cylindrical coordinates ($r, \beta, z$):
\begin{align}
    f(t) &=\int_0^\infty \int_0^{2\pi} p(r,0,t) r\;d\beta \;dr, \\
    &=\pi r_b^2b(t)+2\pi\int_0^{r_b}-\frac{2r^2}{\pi^2(a^2-r^2)} r\;dr + 2\pi\int_0^{r_b} \frac{2aa'}{\pi\sqrt{a^2-r^2}} r\;dr,
\label{sol-f}
\end{align}
for $t \geq 0$, and equation \ref{sol-f} needs to be further evaluated to the radial boundary to obtain a closed-form impact force solution.

\subsection{Contact-line Dynamics} \label{2.2}
\begin{figure}
    \centering
    \begin{subfigure}[b]{\textwidth}
        \caption{}
        \includegraphics[width=0.95\textwidth]{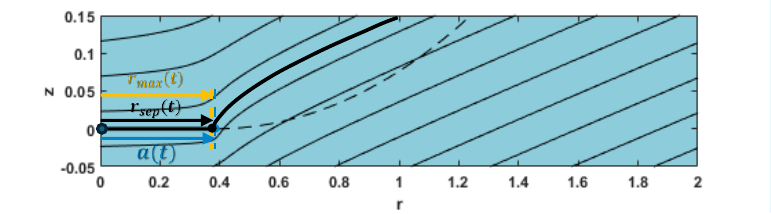}
        \label{figure_corner1}
    \end{subfigure}
    \begin{subfigure}[b]{\textwidth}
        \caption{}
        \includegraphics[width=0.95\textwidth]{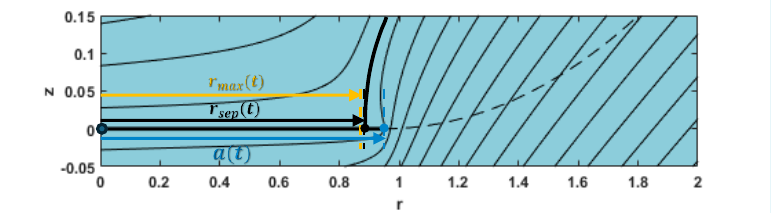}
        \label{figure_corner2}
    \end{subfigure}
    \begin{subfigure}[b]{\textwidth}
        \caption{}
        \includegraphics[width=0.95\textwidth]{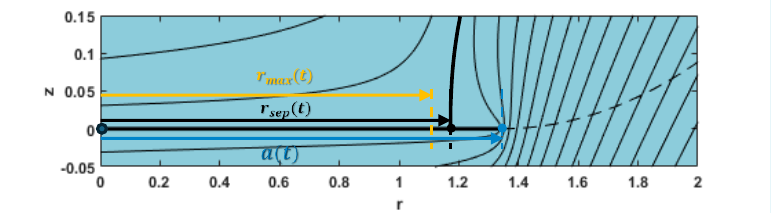}
        \label{figure_corner3}
    \end{subfigure}
    \begin{subfigure}[b]{\textwidth}
        \caption{}
        \includegraphics[width=0.95\textwidth]{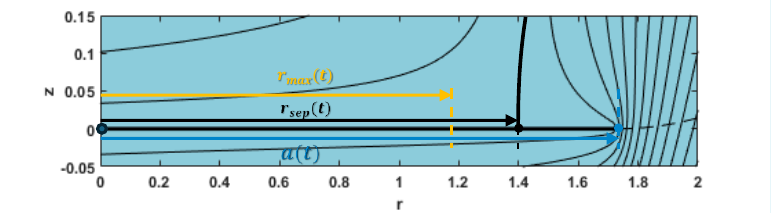}
        \label{figure_corner4}
    \end{subfigure}
    \caption{Positions of the wet radius $a$, the separation point $r_{sep}$, and the pressure-maximum radius $r_{max}$ at times $t=5\times10^{-2}$ (a), $t=3\times10^{-1}$ (b), $t=6\times10^{-1}$ (c), and $t=10^{0}$ (d) are shown overlaid on the streamlines of flows around a flat, rising, expanding disk. The dashed line indicates the droplet interface, analogous to the droplet impact on a solid surface.}
    \label{figure_corner}
\end{figure}

To determine the proper disk expanding speed in equations \ref{sol-u} to \ref{sol-f}, in the context of an analogy to droplet impact, the corresponding time-dependent wet radius for liquid droplet is required as the disk radius $a(t)$. The well-known \cite{Wagner1} condition reveals that the free surface meets the solid disk surface at the turnover point (namely wet radius), and enables derivation of the wet radius $a(t)=\sqrt{3t}$ \citep{Wagner1, Riboux1, Negus1} for `small deadrise angle' droplet impact, such as the case of spherical droplet impact. Furthermore, \cite{Hao1} suggested that, following the principle of Wagner condition, the turnover point shall be the stagnation point in the moving frame of reference at velocity $\bm{a'(t)}$ (with the magnitude $a'(t)$ towards the radial direction) standing at the disk edge, where $x'(t)$ indicates time-derivative $dx/dt$ for general variables. The position of this stagnation point, $r_{sep}$ (denoted as $r_{2}$ in \cite{Hao1}), was named `separation point' for the flow to be distinguished from the conventional `stagnation point' concept of being stationary in the laboratory frame. This led to the radial boundary $r_b=r_{sep}$ of the solutions of equations \ref{sol-u} to \ref{sol-f} \citep{Hao1}:
\begin{equation}
    r_{sep}(t)=\sqrt{\frac{3t}{1+ \frac{16t}{3\pi^2}}}.
\label{r_stag}
\end{equation}
The positions of the separation point at $t=5\times10^{-2}$, $t=3\times10^{-1}$, $t=6\times10^{-1}$ and $t=10^{0}$ are illustrated in figure \ref{figure_corner}.

However, though the derived analytical solutions have so far the longest validations in time, as has been noticed by \cite{Hao1}, solutions break down at $t\approx0.27$, where peak impact force happens, and hence cannot describe the late-time impact processes. Specifically, Hao \emph{et al.} found that the breakdown relates to the improper kinematic boundary condition of the wet radius expanding speed, $a(t)$, which is valid at early times but overestimates the contact-line speed at late times. Deriving an applicable analytical description for the wet-radius dynamics is particularly challenging for intermediate and late times, when the lamella forms. On one hand, \cite{Egger1} and \cite{Garcia2020} provide analytical spreading radii or diameters only for the contact line, encompassing both the droplet and lamella (including the rim). On the other hand, \cite{Sanjay} quantifies the contact line\footnote{\cite{Sanjay} used the word `contact footprint' for $r_f$.} $r_f$ using scaling arguments, maintaining $r_f\sim\sqrt{t}$ across both impact and spreading phases; see also figure 24b of \cite{Hao1} for a sensitivity study showing that moderating the wet-radius dynamics, within the scaling of $a\sim\sqrt{t}$, effectively suppresses exaggerated impact hydrodynamics. Similarly, the Bernoulli constant, $b(t)$, which was found as $b(t\ll1)=1/2$ based on initial flow conditions \citep{Wagner1,Riboux1}, should have decreased with time to $b(t=\textit{O}(1))\rightarrow0$ at late times $t\gtrsim1$ \citep{Hao1}; see more details below. These (partly) explain why the impact loads persist too strongly in the model after the peak impact force instant, $t\approx0.27$. In the present work, we extend the time validity of the solutions of \cite{Hao1} beyond $t\approx0.27$ by introducing new assumptions that specifically address these two limitations.

Before that, we note another important feature of the theoretical framework\textemdash the radius $r_{max}$:
\begin{equation}
  r_{max}(t)=\sqrt{3t-\frac{16t^2}{\pi^2}},
  \label{r_max}
\end{equation}
where the pressure profile of equation \ref{sol-p} has a radial maximum \citep{Hao1}. $r_{max}$ is known as the position of the ring pressure pattern, being widely observed from experiments on impact erosion studies \citep{Field1, Field3, Engel1, Brunton1}. \cite{Hao1} showed the indistinguishable difference between $r_{sep}$ and $r_{max}$ at early times up to $\textit{O}(10^{-1})$ (see their figure 12d), however afterwards, $r_{max}$ quickly converges to the surface centre while $r_{sep}$ monotonically expands (see figure \ref{figure_corner}). Particularly, we note that equation \ref{r_max} is valid for $0 \leq t \leq t_{ter}$, where $t_{ter}=3\pi^2/16$ is the termination time of the ring pressure pattern as the pressure peak moves to the radial origin (\emph{i.e.}, centre-stagnated pressure pattern), as noted by \cite{Hao1}. Below, we evaluate the two radii, $r_{max}$ and $r_{sep}$, in details.

\begin{figure}
    \centering
    \begin{subfigure}[b]{0.45\textwidth}
        \caption{}
        \includegraphics[width=0.95\textwidth]{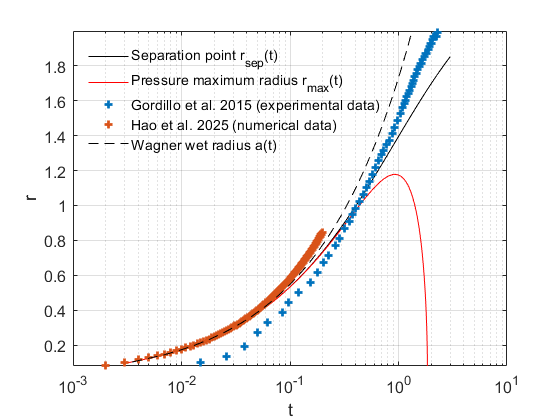}
        \label{figure_evidence1_contactline}
    \end{subfigure}
    \begin{subfigure}[b]{0.45\textwidth}
        \caption{}
        \includegraphics[width=0.95\textwidth]{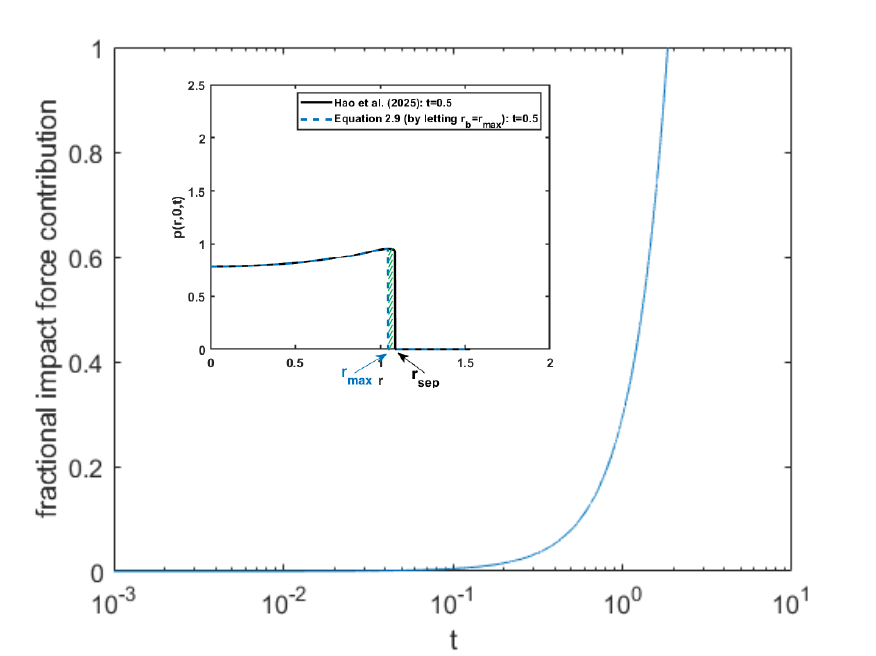}
        \label{figure_evidence1_force}
    \end{subfigure}
    \caption{(a) Temporal evolution of the radial positions of the separation points $r_{sep}$, the pressure maximum radius $r_{max}$, and the Wagner wet radius $a(t)=\sqrt{3t}$. Also superimposed are the contact-line evolutions from experiments \citep{Gordillo2} and numerical simulations \citep{Hao1}, for a liquid droplet with kinematic viscosity $20$ cSt and impact velocity $1.93$ \unit{\meter\per\second} ($Re=212$). (b) Fractional contribution to the impact force from the pressure acting in the radial interval between $r_{max}$ and $r_{sep}$ (\emph{i.e.} the green-shaded region in the inset), normalised by the total impact force within $r_{sep}$. The inset in (b) shows the radial pressure distribution at $t=0.5$.}
    \label{figure_evidence1}
\end{figure}

Figure \ref{figure_evidence1_contactline} shows the temporal evolutions of the radii $r_{max}$ and $r_{sep}$ from early times $t=\textit{O}(10^{-3})$ to late times $t=\textit{O}(1)$ (see also figure 20b of \cite{Hao1} for comparison). The time-axis is shown on a logarithm scale to better resolve the intermediate-time regime, \emph{i.e.} the transition between early and late times. Up to $t=\textit{O}(10^{-1})$, the two radii are indistinguishable, both being inboard of the wet radius $a(t)$, consistent with the observations of \cite{Hao1}. A deviation between $r_{max}$ and $r_{sep}$ emerges for $0.2<t<0.3$ \citep{Hao1}, but remains small until $t\approx1$. At late time, $r_{sep}$ diverges towards its asymptotic limit $3\pi/4$, while $r_{max}$ retreats towards the origin (\emph{i.e.}, $r=0$). Interestingly, this behaviour closely mirrors that of the impact force solution in \cite{Hao1}, which departs from experimental measurements at $t\approx0.27$, and subsequently rises to a quasi-constant value\footnote{This value is not strictly constant and eventually decays to zero at sufficiently long times; see \cite{Hao1}.}, whereas the experimentally measured force decays to zero. This observation motivates the hypothesis that, instead of using $r_{sep}$, adopting a truncated solution boundary $r_b=r_{max}$ can effectively suppress the overestimation of the wet radius from times $\textit{O}(10^{-1})$ onward. In doing so, the contribution of the pressure acting between radii $r_{max}$ and $r_{sep}$ (see the inset of figure \ref{figure_evidence1_force}) is removed, thereby reducing the predicted impact force. This effect is quantified in figure \ref{figure_evidence1_force}, where the difference between the original and truncated solutions is negligible up to $t=\textit{O}(10^{-1})$. At later times, the truncation at $r_{max}$ progressively offsets the overpredicted force and ultimately eliminates it at $t=\textit{O}(1)$, as expected.

For broader validation, figure \ref{figure_evidence1_contactline} also superimposes the contact-line evolution obtained from both experiments \citep{Gordillo2} and numerical simulations \citep{Hao1}. Both datasets exhibit trends similar to that of the separation point $r_{sep}$: they agree with the positions of $r_{max}$ at early times but progressively deviate at later times. It should be noted, however, that these datasets describe the contact-line dynamics, which lie outboard of the wet radius once sheet ejection occurs \citep{Quintero,Gordillo2019}. Therefore, the distinctive late-time behaviour does not contradict the physical fidelity of the present assumption. 

\begin{figure}
    \centering
    \begin{subfigure}[b]{\textwidth}
        \centering
        \caption{}
        \includegraphics[width=0.6\textwidth]{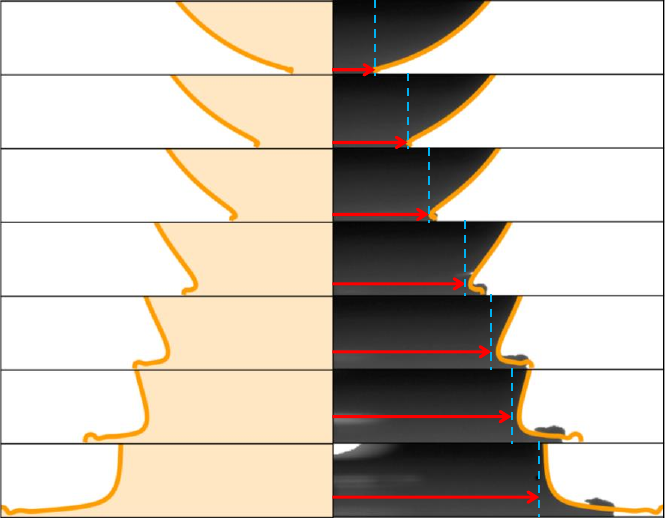}
        \label{figure_evidence31}
    \end{subfigure}
    \begin{subfigure}[b]{\textwidth}
        \centering
        \caption{}
        \includegraphics[width=0.85\textwidth]{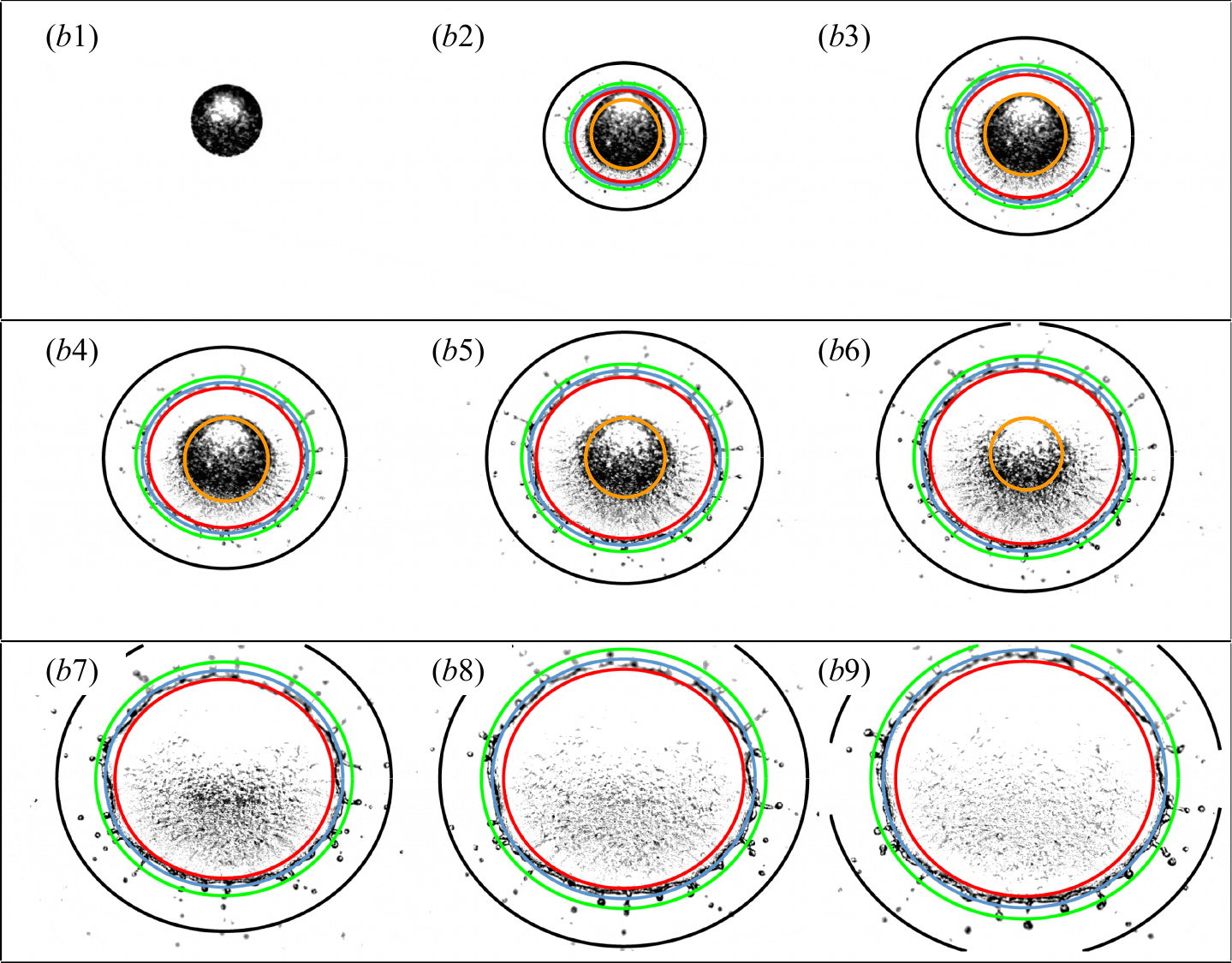}
        \label{figure_evidence32}
    \end{subfigure}
    \caption{(a) Comparison between the predicted radius $r_{max}$ (red arrow and dashed line) and experimentally observed contact-line positions from \cite{Riboux2016} for an ethanol droplet of radius $R_0=1.03$ \unit{\milli\meter}. The dimensionless times (from top to bottom) are $t=0.016$, $0.049$, $0.082$, $0.164$, $0.246$, $0.328$, and $0.491$. The left column shows the corresponding numerical results in \cite{Riboux2016}. (b) Comparison between the predicted radius $r_{max}$ (orange line) and experimentally observed droplet morphologies from \cite{Quintero} for a water droplet of radius $R_0=1.53$ \unit{\milli\meter}. The dimensionless times corresponding to labels $b$1 to $b$9 are $t=0$, $0.4$, $0.7$, $0.9$, $1.2$, $1.4$, $1.9$, $2.4$, and $2.9$. See \cite{Quintero} for other coloured lines.}
    \label{figure_evidence3}
\end{figure}

For a more informative comparison, figure \ref{figure_evidence3} overlays the predicted positions of $r_{max}$ onto experimentally observed droplet morphologies. Figure \ref{figure_evidence32} shows a side view of the contact-line dynamics from early to intermediate times. Prior to sheet ejection (frames 1-2), $r_{max}$ aligns precisely with the wet radius (or equivalently the contact line). Once the lamella is ejected (frames 3-7), $r_{max}$ becomes progressively inboard of the wet radius, denoted as the turning point of the droplet interface curvature \citep{Riboux1,Philippi1}. However, the separation between these two locations remains small until the final frame, where the interface curvature becomes difficult to identify. At intermediate to late times, figure \ref{figure_evidence31} presents a perpendicular view of the impacting droplet and its spreading lamella. Before $r_{max}$ retreats towards the origin (frames $b$1 to $b$4), it closely follows the shape of the bulk of the droplet. Thereafter (frames $b$4 to $b$6), $r_{max}$ retreats together with the diminishing droplet bulk, eventually collapsing to the origin (frames $b$7 to $b$9) as the bulk liquid feeds into the spreading lamella. During this very late stage, changes in the spreading diameter are subtle as the droplet approaches its maximum spreading, and the associated impact loads are therefore expected to be weak \citep{Cheng}. Overall, these comparisons demonstrate that $r_{max}$ provides a meaningful characterisation of the full hydrodynamic evolution, capturing the early-time impact, the intermediate-time transition, and the decaying late-time spreading regimes.

However, it is important to note that the above modification has inherent physical limitations. Truncating the flow solution at $r_{max}$ effectively neglects the outer portions of the spreading droplet (see the inset of figure \ref{figure_evidence1_force}), which do contribute to the pressure and contact area, albeit with smaller magnitudes. As a consequence, several phenomena associated with contact-line dynamics \citep{Riboux1,Gordillo2019}\textemdash including sheet ejection, dewetting and splashing\textemdash as well as the influence of ambient air on the impact dynamics \citep{Sprittles}, cannot be captured. Nevertheless, as demonstrated above, this modification yields clear benefit from a hydrodynamic perspective, notably a converged force profile (see section \ref{2.3}) and, consequently, a finite impact impulse on the fluid-solid system. This provides a sound basis for further use of the analytical solution in the present study (see section \ref{3}) and other in related engineering applications. Moreover, for practical considerations in impact-erosion studies, the primary, if not sole, quantity of interest is at $r_{max}$ where the impact pressure attains its maximum. The proposed modification therefore does not sacrifice quantities of relevance for engineering erosion applications. In view of this trade-off, the present modification is benign for impact-erosion studies, although it may not be appropriate for other impact scenarios.

\begin{figure}
    \centering
    \includegraphics[width=0.45\linewidth]{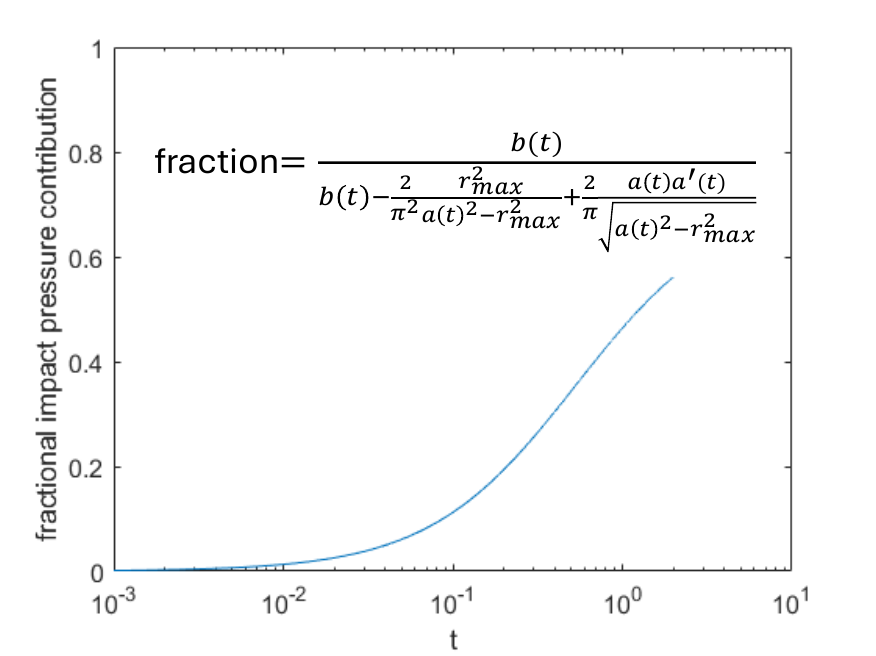}
    \caption{Fractional contribution of the Bernoulli constant $b(t)=1/2$ to the impact pressure, normalised by the peak impact pressure $p(r_{max},0,t)$, as a function of time $t$. The expression used to compute this fraction is shown in the figure, with $a(t)=\sqrt{3t}$.}
    \label{figure_evidence2}
\end{figure}

Regarding the time-dependent Bernoulli constant $b(t)$, it can be derived as $1/2$ from the far-field condition of equation \ref{BC3} in the inviscid framework of potential-flow theory. However, at late times $t\approx1$, in the context of droplet impact\textemdash where the flow field is not infinite, this far-field condition should instead be corrected by the flow velocity at a height of $\textit{O}(1)$ above the impact surface, which would drive the Bernoulli constant towards zero. At present, there is no known analytical expression for a time-dependent Bernoulli constant in the droplet impact problem that describes the transition between these two limits; a detailed discussion can be found in \cite{Hao1}. In this sense, retaining the constant value $b(t)=1/2$ leads to an overprediction of the impact pressure as time progresses. Figure \ref{figure_evidence2} quantifies the fractional contribution of the Bernoulli constant term $b(t)=1/2$, normalised by the peak impact pressure $p(r_{max},0,t)$, as a function of time $t$. A behaviour similar to that shown in figure \ref{figure_evidence1_force} is observed: the Bernoulli constant contribution is, notably, negligible at early times but progressively dominates the impact pressure magnitude at $t=\textit{O}(1)$, consistent with the findings of \cite{Hao1}. Motivated by this similar pattern, we introduce a secondary modelling modification in which the contribution of the Bernoulli constant is removed to suppress the late-time overestimation of the hydrodynamic response from times $\textit{O}(10^{-1})$ onward, while having negligible influence on early-time hydrodynamics. Accordingly, we set $b=0$ throughout the impact duration, in line with previous studies \citep{Philippi1,Negus1}.

This choice leads to a slight underestimation of the impact loads at intermediate times $t=\textit{O}(10^{-2})$ (and small $t=\textit{O}(10^{-1})$), as shown in figure \ref{figure_evidence2} (see also section \ref{2.3}), where the contribution of the Bernoulli constant, although small, is not negligible. Nevertheless, as demonstrated above, these modifications (combine to) extend the analytical solutions to longer times, providing an approximate analytical closure for the impact and spreading phases. Although these modifications are not derived purely from first-principles physics (experimental and numerical evidence is provided in later sections), they yield a tractable analytical description of droplet impact that can be directly applied to engineering impact erosion studies, at the cost of only a minor loss of physical fidelity.

Finally, we reiterate that the purpose of the present study is not to derive a complete analytical solution describing the entire droplet-impact process (see literature review in section \ref{1} for the four impact regimes), but rather to develop a tractable analytical model suitable for engineering applications. In this sense, the modified analytical solution is appropriate for the intended purpose, despite not capturing the (rather interesting) phases of maximum spreading, and splashing, pinning, retraction and jump-off. This limitation is justified by the fact that the impact pressure and force are weak, or even negligible, during these late-time regimes on non-hydrophobic surfaces \citep{Egger1,Sanjay,Garcia2020,Cheng}.

\subsection{A Modified Solution to Surface Loadings upon Liquid Droplet Impact} \label{2.3}
With the Wagner's \citep{Wagner1} wet radius $a(t)=\sqrt{3t}$ for liquid droplet and the above considerations, namely to pose the spatial boundary at $r_{max}$ and set the Bernoulli constant term $b(t)$ to zero, we modify the analytical solutions in section \ref{2.1.2}, on the solid disk surface ($z=0$), as shown below:
\begin{equation}
  \bm{u}(r,0,t) = (\frac{2r}{\pi\sqrt{3t-r^2}}) \bm{\underline{\hat{r}}} + \bm{\underline{\hat{z}}} \;\;\;\;\;\;\;\; \text{for $r < r_{max}(t)$},
\label{sol-u2}
\end{equation}
and:
\begin{equation}
  \frac{\partial \phi}{\partial t}(r,0,t) =- \frac{3}{\pi \sqrt{3t-r^2}} \;\;\;\;\;\;\;\; \text{for $r < r_{max}(t)$},
\label{sol-dphidt2}
\end{equation}
with $r_{max}(t)$ from equation \ref{r_max}. These lead to the analytical loadings on the solid surface upon impact as:
\begin{equation}
  p(r,0,t) = - \frac{2}{\pi^2} \frac{r^2}{3t-r^2} + \frac{3}{\pi \sqrt{3t-r^2}} \;\;\;\;\;\;\;\; \text{for  $r <  r_{max}(t)$} ,
\label{sol-p2}
\end{equation}
and:
\begin{align}
    f(t) &=2\pi\int_0^{r_{max}}-\frac{2r^2}{\pi^2(3t-r^2)} r\;dr + 2\pi\int_0^{r_{max}} \frac{3}{\pi\sqrt{3t-r^2}} r\;dr, \nonumber\\
          &=\frac{2}{\pi}\left(r_{max}^2 log(m)-( 3t (log(3t) -1) - m(log(m)-1))\right) + 6(\sqrt{3t}-\sqrt{m}),
\label{sol-f2}
\end{align}
for $t \geq 0$, where $m(t)=3t-r_{max}^2(t)$. The modified solutions of equations \ref{sol-u2} to \ref{sol-f2} extend the solutions from \cite{Hao1} beyond $t\approx0.27$ to a full-duration analytic approximation (see below) by implementing new assumptions discussed in section \ref{2.2}.

\begin{figure}
    \centering
    \begin{subfigure}[b]{0.45\textwidth}
        \caption{}
        \includegraphics[width=0.95\textwidth]{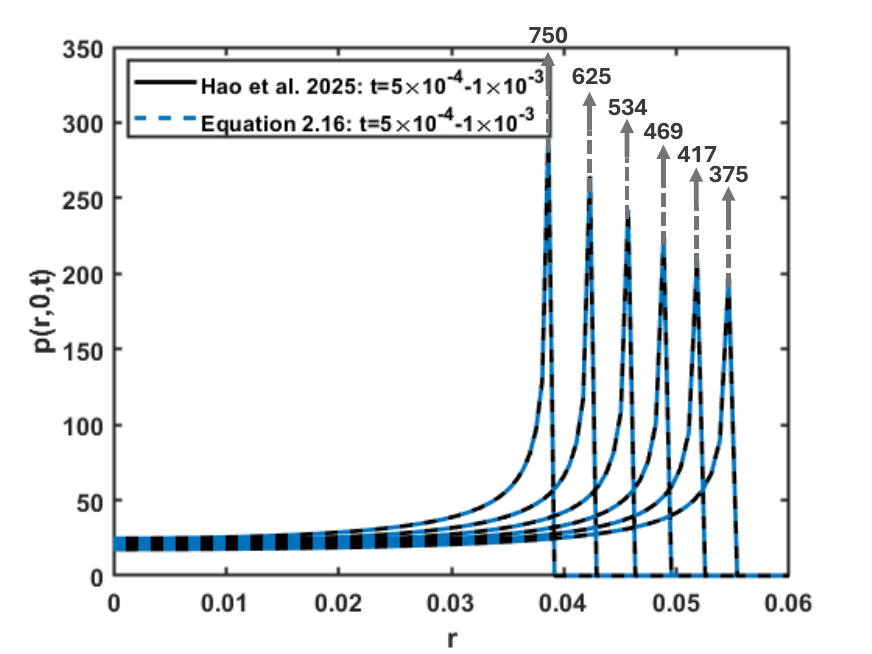}
        \label{figure_pressure1}
    \end{subfigure}
    \begin{subfigure}[b]{0.45\textwidth}
        \caption{}
        \includegraphics[width=0.95\textwidth]{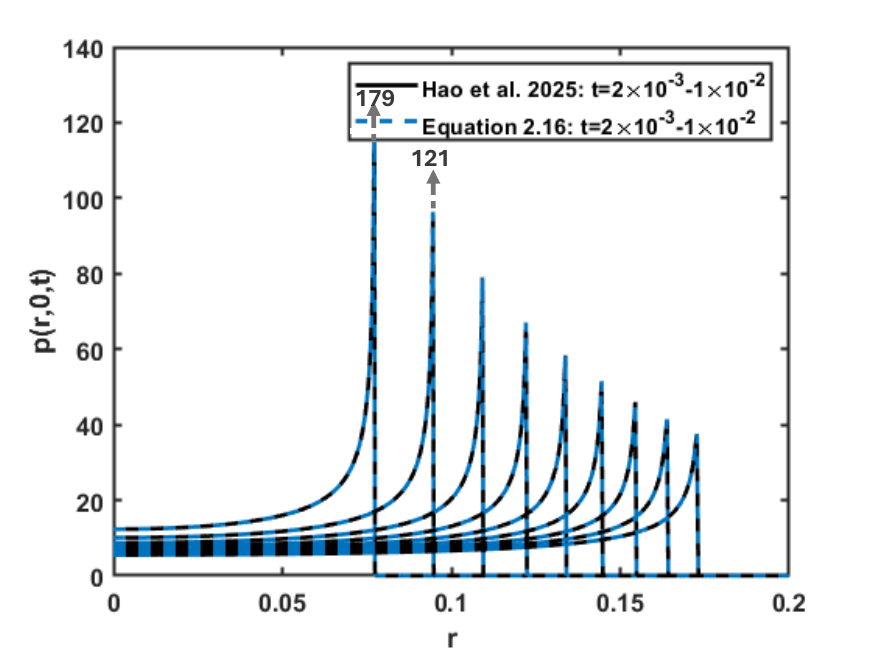}
        \label{figure_pressure2}
    \end{subfigure}
    \begin{subfigure}[b]{0.45\textwidth}
        \caption{}
        \includegraphics[width=0.95\textwidth]{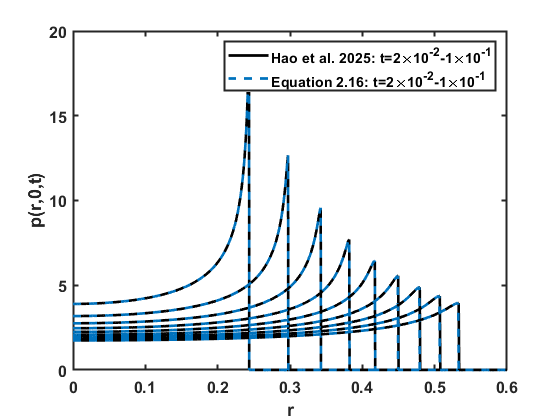}
        \label{figure_pressure3}
    \end{subfigure}
    \begin{subfigure}[b]{0.45\textwidth}
        \caption{}
        \includegraphics[width=0.95\textwidth]{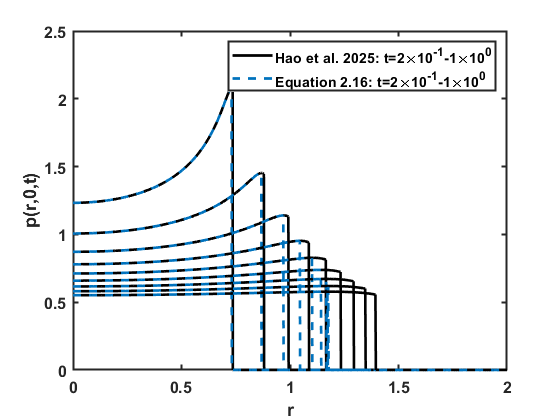}
        \label{figure_pressure4}
    \end{subfigure}
    \caption{Analytical radial pressure profiles from equation \ref{sol-p2} are shown for four time decades: $t=\textit{O}(10^{-4})$ (a), $t=\textit{O}(10^{-3})$ (b), $t=\textit{O}(10^{-2})$ (c), and $t=\textit{O}(10^{-1})$ (d), compared with the analytical solutions of \cite{Hao1}. In panels (a) and (b), for $t=5\times10^{-4}$ to $t=3\times10^{-3}$, the peak values (identical for both analytical solutions) exeed the plot limits, indicated by arrows pointing to each peak, but the horizontal plotting resolution is insufficient to fully capture them.}
    \label{figure_pressure}
\end{figure}

\begin{figure}
    \centering
    \includegraphics[width=1\linewidth]{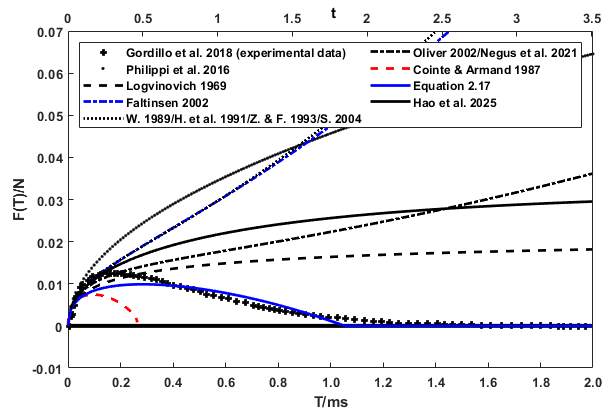}
    \caption{Time-series of the analytical impact force from equation \ref{sol-f2} on the solid surface, compared with experimental data from \cite{Gordillo2} (scatter points). Impact parameters are: $U_0 = \qty{1.93}{\meter\per\second}$ at normal incidence, kinematic viscosity $20$ cSt, $Re = 106$, and $We = 56$. Additional analytical solutions from the literature are also shown for comparison. The secondary $x$-axis at the top indicates the dimensionless time $t$.}
    \label{figure_force}
\end{figure}

Impact pressure of equation \ref{sol-p2} is shown in figure \ref{figure_pressure} for four decades in time $\textit{O}(10^{-4})$ - $\textit{O}(10^{-1})$ to compare the original solutions of \cite{Hao1}. Readers are referred to their figure $19$ for comparison. As we can see, indistinguishable difference is observed for times before $\textit{O}(10^{-1})$, after which the modified solution starts to have a lesser radial boundary. The corresponding impact force on the surface is presented in figure \ref{figure_force}. For validation purpose, impact parameters are chosen from the experiment in \cite{Gordillo2}. As can be seen, although the modified analytical solution does not capture the full physics of the impact process, it does avoid the early breakdown of the prior solution at $t\approx0.27$ and can be run until the droplet has essentially finished spreading, including a reasonable, though not exact, prediction of the tail of the time-series. The observation agrees with the literature that impact loadings are dominated by the impact and spreading phases \citep{Egger1, Gordillo2, Cheng, Sanjay, Garcia2020}. This provides a tractable analytical solution for droplet impact in engineering applications. Solutions of equations \ref{sol-p2} and \ref{sol-f2} of the impact loading on the solid surface are needed for the impact erosion study of the solid material, and in the subsequent sections, we will report how the modified analytical solution can be useful in such purposes.

As a final note before applying the analytical flow solution to practical impact problems, we comment on the range of applicability of the modified analytical solution. The analytical solution was originally derived by \cite{Hao1} and subsequently modified in the present study within the framework of inviscid potential flow theory, which neglects both viscous and capillary effects. As noticed in the literature, although viscosity may be of secondary importance at early times \citep{Cheng}, viscous damping plays a significant role in the late-time decay of the impact force during the spreading regime \citep{Roisman2,Roisman1,Sanjay3,Garcia2020}, even in flow regimes that are often considered inviscid \citep{Sanjay}. Similarly, capillary effects are known to be important in the late-time regimes of droplet impact \citep{Egger1,Lagubeau,Garcia2020,Sanjay}, particularly near the maximum spreading diameter \citep{Cheng}\footnote{However, impact force is very weak near the maximum spreading \citep{Cheng}, as demonstrated above in section \ref{2.2}.}. This is in addition to other capillary-related flow phenomena, including bubble entrapment \citep{Thoroddsen2005,Josserand, Garcia2022} and gas-film cushioning \citep{Gordillo2022,Sprittles,Garcia2024,Gabbard}. Accordingly, although the modifications introduced in section \ref{2.2} extend the potential flow solution to provide an approximate analytical closure for both the impact and spreading phases, we conservatively restrict the applicability of the modified analytical solution to inertia-dominated regime of droplet impact characterised by large Reynolds and Weber numbers.

\section{Numerical Models} \label{3}
\subsection{Physical Model} \label{3.1}
\begin{figure}
    \centering
    \includegraphics[width=0.8\linewidth]{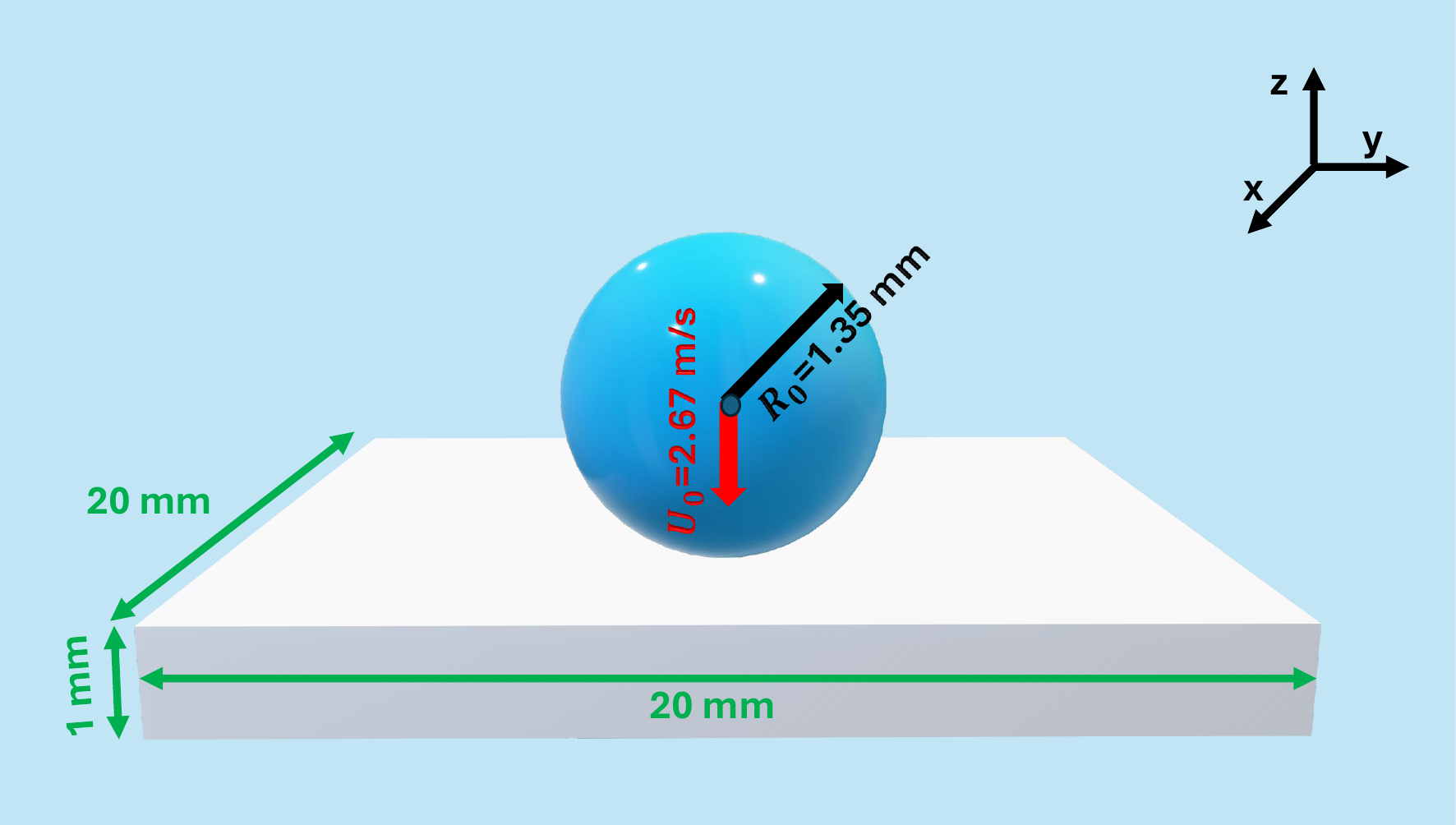}
    \caption{Three-dimensional SPH simulation domain showing a water droplet of radius $R_0=\qty{1.35}{\milli\meter}$ impacting normally on an aluminium alloy plate at $U_0=\qty{2.67}{\meter\per\second}$. The aluminium plate measures \qty{20}{\milli\meter} in length and width, and \qty{1}{\milli\meter} in thickness. The figure is for illustration purpose and not to scale.}
    \label{figure_framework}
\end{figure}

In this section we apply the derived analytical solutions from section \ref{2.3} to a water droplet impact study on solid material surfaces in the context of material impact erosion. We will compare results with the conventional numerical methods used in the field. To our knowledge, \cite{XJ1} is the only available experimental work on liquid droplet impacting solid materials (with high stiffness) with experimental data from material measurements. Thereby, for validation purposes, the simulated impact conditions and parameters in this section will model the laboratory test conditions in \cite{XJ1}. The numerical simulation domain of the present study is shown in figure \ref{figure_framework}. As in the literature, water droplet of radius $R_0=\qty{1.35}{\milli\meter}$, with density $\rho_l=\qty{995.8}{\kilo\gram\per\meter\cubed}$ and viscosity $\mu_l=\qty{0.001}{\pascal\second}$, impacts an aluminium alloy plate at a speed of $U_0=\qty{2.67}{\meter\per\second}$ at normal impact angle. We note the Reynolds number $Re=\rho_lU_0R_0/\mu_l=7355$, Weber number $We=\rho_lU_0^2R_0/\sigma=261$ (with surface tension $\sigma=\qty{0.073}{\newton\per\meter}$), and the Froude number $Fr=U_0/\sqrt{gR_0}=23.3$ (with gravitational acceleration $g=\qty{9.81}{\meter\per\second\squared}$) of the simulated multi-phase problem, which are high enough to neglect the surface tension and gravity. Therefore, the liquid phase lies within the range of applicability of the analytical framework, and, where numerical simulations are employed, only a water droplet is simulated, with the ambient air neglected. The aluminium alloy plate is of the size $\qty{20}{\milli\meter}$(length)$\times\qty{20}{\milli\meter}$(width)$\times\qty{1}{\milli\meter}$(height), and the material properties are summarised in table \ref{table_mat}. 

\subsection{Numerical Treatments} \label{3.2}
The conventional method for material analysis is the finite-element (FE) method \citep{XJ1, Keegan1, Verma1}, and particularly for such fluid-structure interaction problems, two techniques, namely the coupled Eulerian-Lagrangian (CEL) and smoothed particle hydrodynamics (SPH) are widely used. The former simulates the material phase using the conventional Lagrangian FE approach, while for the liquid phase, an Eulerian approach is used to `assign' materials to the mesh grids at each simulation step. Similarly in SPH, the liquid phase is represented by a cloud of mass particles that contain the position and mass information of the fluid elements. Since the two methods provide similar performance \citep{Doagou1} and Zhang \emph{et al.} used SPH in their numerical simulation \citep{XJ1}, here we simulate the impact problem using SPH. 

\begin{table}
    \centering
    \caption{Material properties of the aluminium plate used in FE simulation, as reported in \citep{XJ1}.}
    \begin{tabular}{cccc}
        \hline
        \bf{Material} & \bf{Young's Modulus $E$} & \bf{Density $\rho_s$} & \bf{Poisson's Ratio $\nu$}\\
        \hline
        Aluminium Alloy & \qty{70000}{\mega\pascal} & \qty{2820}{\kilo\gram\per\meter\cubed} & $0.3$\\
        \hline
    \end{tabular}
    \label{table_mat}
\end{table}

\begin{figure}
    \centering
    \includegraphics[width=0.8\linewidth]{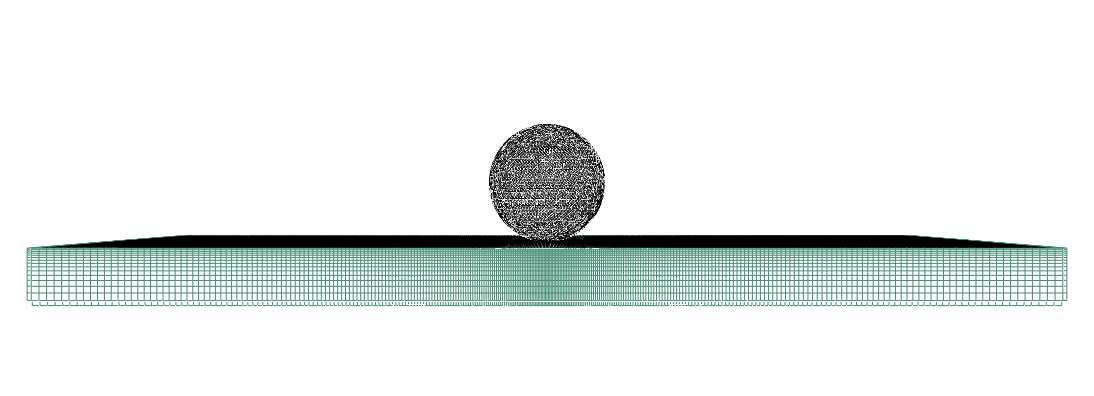}
    \caption{Illustration of the SPH mesh used in the present study for the droplet (black) and aluminium plate (green). This mesh corresponds to the $SPH5$ simulation (and the aluminium plate in $ANCM5$) listed in table \ref{table_mesh}.}
    \label{figure_mesh}
\end{figure}

The three-dimensional SPH model is implemented in the commercially available software ABAQUS\raisebox{1ex}{\scriptsize \textregistered}, version $2024$ Research \citep{ABA}. The water droplet is modelled in ABAQUS with the continuum particle elements (PC3D). The aluminium alloy plate is modelled using eight-node brick elements with reduced integration (C3D8R) with a layer of one-way infinite elements (CIN3N8) on the four vertical sides and the bottom horizontal surface to eliminate wave reflections due to the finite geometry. The four sides and the bottom surface are fixed using encaster boundary conditions to be consistent with the numerical model in the literature \citep{XJ1}. For the finest mesh used in the present study, the plate is bias-meshed with a finer mesh of size \qty{27}{\micro\meter} (equivalent to \qty{1}{\percent} of droplet diameter) towards the centre of the impact surface and coarser further away (\qty{150}{\micro\meter}, see figure \ref{figure_mesh}). The size of the fine mesh is selected to align with the guidance in the community of FE simulation on droplet impact \citep{Doagou1} that the finest mesh shall be on the order of \textit{O}(\qty{1}{\percent}) of the droplet diameter. The droplet geometry is meshed with a uniform fine mesh of size \qty{80}{\micro\meter}. Due to the large deformation of the simulated problem, finer meshes would require an extremely large number of increments to converge and hence we try to avoid. The total number of elements (including the one-way infinite layers) is $\textit{O}(10^6)$. Total simulation time is \qty{2}{\milli\second} (corresponding to dimensionless time \qty{4.0}) at a timestep of \qty{10}{\micro\second}. The timestep is selected so that the peak force shall be captured in no less than ten steps \citep{Doagou1}, and our sensitivity check indeed showed no difference in results at finer timesteps. A typical simulation took roughly $10^3$ CPU hours to solve on a AMD Ryzen Threadripper PRO 5975WX\raisebox{1ex}{\scriptsize TM} processor with \qty{270}{\giga\byte} of RAM.

\subsection{Analytical-Numerical Coupled Model} \label{3.2}
In addition to the SPH simulation described above, we propose an alternative analytical-numerical coupled model (ANCM) by coupling the computational FE material analysis with analytical solutions of liquid droplet impact. 

\begin{figure}
    \centering
    \includegraphics[width=\linewidth]{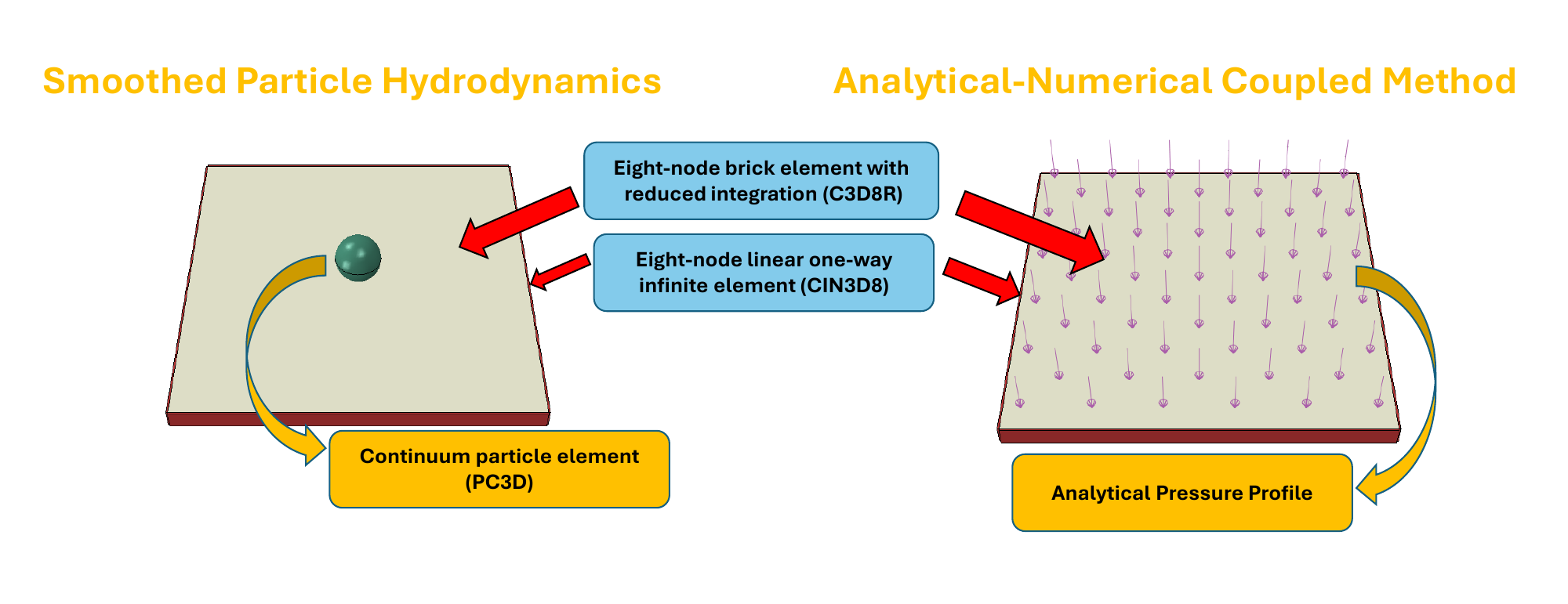}
    \caption{Comparison between the conventional numerical SPH method and the proposed ANCM method. Shared FE elements (with ABAQUS element codes) are highlighted in blue boxes, while differences are indicated in yellow boxes.}
    \label{figure_SPHvsVDL}
\end{figure}

Upon droplet impact, all the information from the fluid dynamics of the droplet is transferred to the solid material through the contact surface. Therefore, with known analytical solutions on the contact surface, we replace the need of liquid phase simulation in FE analysis. The one-way FSI coupling framework is developed by posing the analytical pressure solution of equation \ref{sol-p2} as a time- and spatial-dependent loading on the surface of the solid material to model the impact loading of a liquid droplet. This is achieved in ABAQUS through the VDLOAD user subroutine for explicit dynamic analysis. Figure \ref{figure_SPHvsVDL} illustrates the difference between the SPH and the ANCM models. The same boundary conditions and mesh discretisation were used on the aluminium plate for both SPH and ANCM.

\begin{table}
 \centering
  \caption{Five FE mesh resolutions used for SPH and ANCM simulations in the present study. Element sizes are presented as percentages of $D_0$. For the aluminium plate, the element sizes refer to the finest mesh at the plate's impact surface centre.}
  \begin{tabular}{c|c|c|c}
    \hline
    Cases & Droplet & Aluminium Plate & Total Element Number \\
    \hline\hline
    \textit{SPH}$1$ & \qty{7.4}{\percent} & \qty{9.3}{\percent} & $40\times10^{3}$\\
    \hline
    \textit{SPH}$2$ & \qty{4.2}{\percent} & \qty{4.2}{\percent} & $287\times10^{3}$\\
    \hline
    \textit{SPH}$3$ & \qty{3.7}{\percent} & \qty{3.7}{\percent} & $490\times10^{3}$\\
    \hline
    \textit{SPH}$4$ & \qty{3.0}{\percent} & \qty{3.0}{\percent} & $960\times10^{3}$\\
    \hline
    \textit{SPH}$5$ & \qty{3.0}{\percent} & \qty{1.0}{\percent} & $1247\times10^{3}$\\
    \hline\hline
    \textit{ANCM}$1$ & N/A & \qty{9.3}{\percent} & $33\times10^{3}$\\
    \hline
    \textit{ANCM}$2$ & N/A & \qty{4.2}{\percent} & $256\times10^{3}$\\
    \hline
    \textit{ANCM}$3$ & N/A & \qty{3.7}{\percent} & $458\times10^{3}$\\
    \hline
    \textit{ANCM}$4$ & N/A & \qty{3.0}{\percent} & $888\times10^{3}$\\
    \hline
    \textit{ANCM}$5$ & N/A & \qty{1.0}{\percent} & $1175\times10^{3}$\\
    \hline
  \end{tabular}
  \label{table_mesh}
\end{table}

\section{Results} \label{4}
\subsection{Numerical Models: Validations and Grid-independence Test} \label{4.1}

\begin{figure}
    \centering
    \begin{subfigure}[b]{0.8\textwidth}
        \caption{}
        \includegraphics[width=\textwidth]{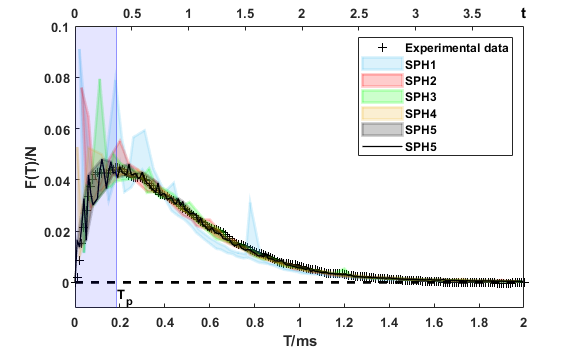}
        \label{figure_num_force1}
    \end{subfigure}
    \hfill
    \begin{subfigure}[b]{0.8\textwidth}
        \caption{}
        \includegraphics[width=\textwidth]{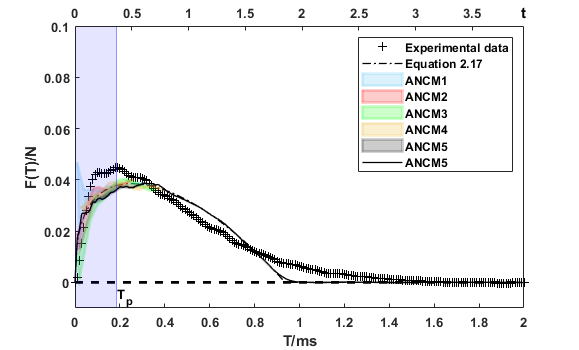}
        \label{figure_num_force2}
    \end{subfigure}
    \caption{Simulated impact force on the aluminium plate surface from SPH (a) and ANCM (b) models, compared with experimental data \citep{XJ1}. Impact parameters are as shown in figure \ref{figure_framework}. Shaded areas indicate the range of spurious oscillations for each method across five FE mesh resolutions listed in table \ref{table_mesh}. The peak impact force from the experimental data occurs at $T_p=\qty{0.19}{\milli\second}$ (dimensionless time $t_p=0.38$).}
    \label{figure_num_force}
\end{figure}

Figure \ref{figure_num_force} compares the impact force acting on the aluminium plate predicted by the numerical simulations with the experimental measurements reported in \cite{XJ1}. To assess numerical convergence, both the SPH and ANCM models were evaluated using five different mesh resolutions, as summarised in table \ref{table_mesh}. For the SPH simulations, the envelopes of the numerically induced spurious oscillations at each resolution are superimposed in figure \ref{figure_num_force1}. These oscillations arise from the discrete particle dynamics inherent to the SPH formulation. As shown, the oscillation amplitude is pronounced at coarse resolutions but decreases with mesh refinement, eventually converging at resolutions \textit{SPH}$4$ and \textit{SPH}$5$. At these resolutions, the SPH results agree well with the experimental data. In particular, the converged SPH simulations reproduce the post-peak force decay with high fidelity, including the `tail' of the time series following the peak force instant $T_p=\qty{0.19}{\milli\second}$.

\begin{figure}
    \centering
    \includegraphics[width=0.8\linewidth]{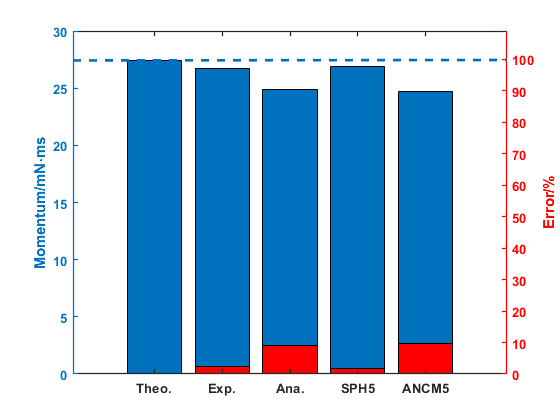}
    \caption{Comparison of the theoretical initial vertical momentum of the droplet with the impact momentum on the aluminium surface calculated from experimental data \citep{XJ1}, the analytical solution of the present study, and numerical predictions from SPH and ANCM simulations. Relative errors with respect to the theoretical momentum are superimposed in red.}
    \label{figure_momentum}
\end{figure}

In contrast, the ANCM results shown in figure \ref{figure_num_force2} exhibit markedly improved grid-independence, characterised by substantially reduced oscillations at comparable mesh resolutions and much faster convergence, achieved already at \textit{ANCM}$2$. However, the ANCM predictions do not fully coincide with the experimental measurements, as the analytical impact force given by equation \ref{sol-p2} (also shown in in figure \ref{figure_num_force2}) deviates slightly from the experimental force history. Notably, the ANCM simulation results agree exactly with the analytical impact force, as expected. It should be emphasised, however, that this agreement is not universal for all solid materials. In cases involving highly deformable solids, discrepancies may arise because the ANCM simulations account for force transmission to a deforming solid surface during the FE analysis, whereas the analytical solution assumes impact onto a rigid boundary. To further quantify the differences between the analytical and numerical predictions, figure \ref{figure_momentum} compares the impact momentum, obtained by time integration of the impact force, with the theoretical initial droplet momentum
\begin{equation}
    M_{Theo}=\frac{4}{3} \rho \pi R_0^3 U_0.
\end{equation}
As shown in figure \ref{figure_momentum}, the momentum predicted by ANCM closely follows the analytical solution, and both approximate the theoretical value within an acceptable margin of accuracy. In comparison, the experimental measurements and the conventional SPH simulations yield slightly more accurate estimates of the initial momentum.

\begin{figure}
    \centering
    \begin{subfigure}[b]{0.495\textwidth}
        \caption{}
        \includegraphics[width=0.95\textwidth]{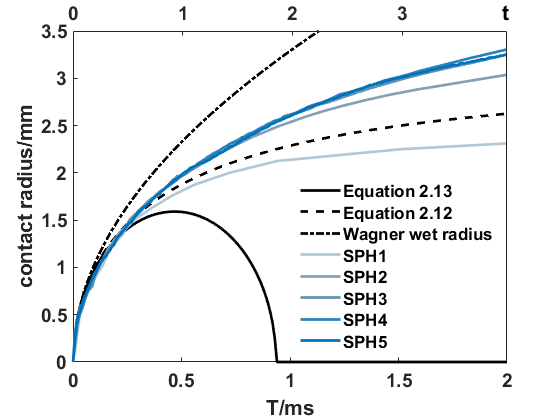}
        \label{figure_num_converge1}
    \end{subfigure}
    \hfill
    \begin{subfigure}[b]{0.495\textwidth}
        \caption{}
        \includegraphics[width=0.95\textwidth]{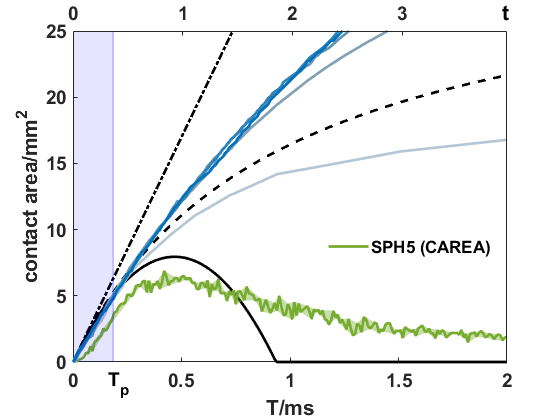}
        \label{figure_num_converge2}
    \end{subfigure}
    \caption{Grid-independence test of the SPH model showing (a) the numerical contact radius and (b) the corresponding calculated contact area. Also shown are the analytical pressure-maximum radius $r_{max}$ (equation \ref{r_max}), separation point $r_{sep}$ (equation \ref{r_stag}), and Wagner wet radius $a$. In (b), the converged ABAQUS output of the total contact area (CAREA) from the $SPH5$ simulation is also superimposed.}
    \label{figure_num_converge}
\end{figure}

Figure \ref{figure_num_converge1} presents the numerical contact radius obtained from SPH simulations using five different mesh resolutions. As the mesh is refined, the predicted contact radius increases and ultimately converges to a stable profile at \textit{SPH}$3$. Also shown are the Wagner wet radius $a$, the separation point $r_{sep}$ given by equation \ref{r_stag}, and pressure-maximum radius $r_{max}$ from equation \ref{r_max}. Consistent with the observations of \cite{Hao1}, the Wagner wet radius overpredicts the numerical contact line. Meanwhile, $r_{sep}$ slightly underestimates the converged SPH contact radius. By contrast, $r_{max}$ predicts a substantially smaller radius than both the analytical and numerical contact lines, and decreases to zero after $t_{ter}\approx1.85$.

Figure \ref{figure_num_converge2} shows the evolution of the contact area, defined as $\pi$ times the square of the contact radius shown in figure \ref{figure_num_converge1}. As the mesh resolution increases, the SPH simulation results exhibit a similar convergence trend to that observed for the contact radius. Our focus here, however, is on the behaviour of the analytical predictions. The contact area computed from $r_{max}$ (equation \ref{r_max}) coincides with that obtained from $r_{sep}$ (equation \ref{r_stag}) up to the peak force instant $T_p$. Beyond this point, the two diverge, with the contact area associated with $r_{sep}$ continuing to increase monotonically. Also shown is the ABAQUS output of the total contact area (CAREA) from the SPH simulations, defined as the sum of all facets experiencing non-zero contact forces. CAREA differs from the geometrically calculated contact area, as it represents the effective force-transmitting area between the liquid and solid phases. The discrepancy between the CAREA and the calculated contact area reflects regions of the spreading or splashing lamella that are in contact geometrically but do not contribute to force transfer. After an initial increase, CAREA decreases and eventually vanishes as the impact transitions from predominantly normal loading to to lateral spreading. Notably, CAREA closely follows the contact area predicted by $r_{max}$, with both reaching a maximum at approximately $t\approx0.5$ and subsequently decaying to zero. This agreement supports the modified analytical solution proposed in section \ref{2.3}, indicating that $r_{max}$ provides a more representative measure of the effective impact loading area, particularly beyond $T_p$, when `touching' contacts emerge due to droplet spreading or splashing \citep{Gordillo2019,Garcia2020}.

\subsection{Material Analysis: Qualitative Comparisons}\label{4.2}
Following the mesh-convergence studies, all SPH results reported in this section correspond to the \textit{SPH}$5$ simulation, while all ANCM results are taken from \textit{ANCM}$5$.

\begin{figure}
    \centering
    \includegraphics[width=\linewidth]{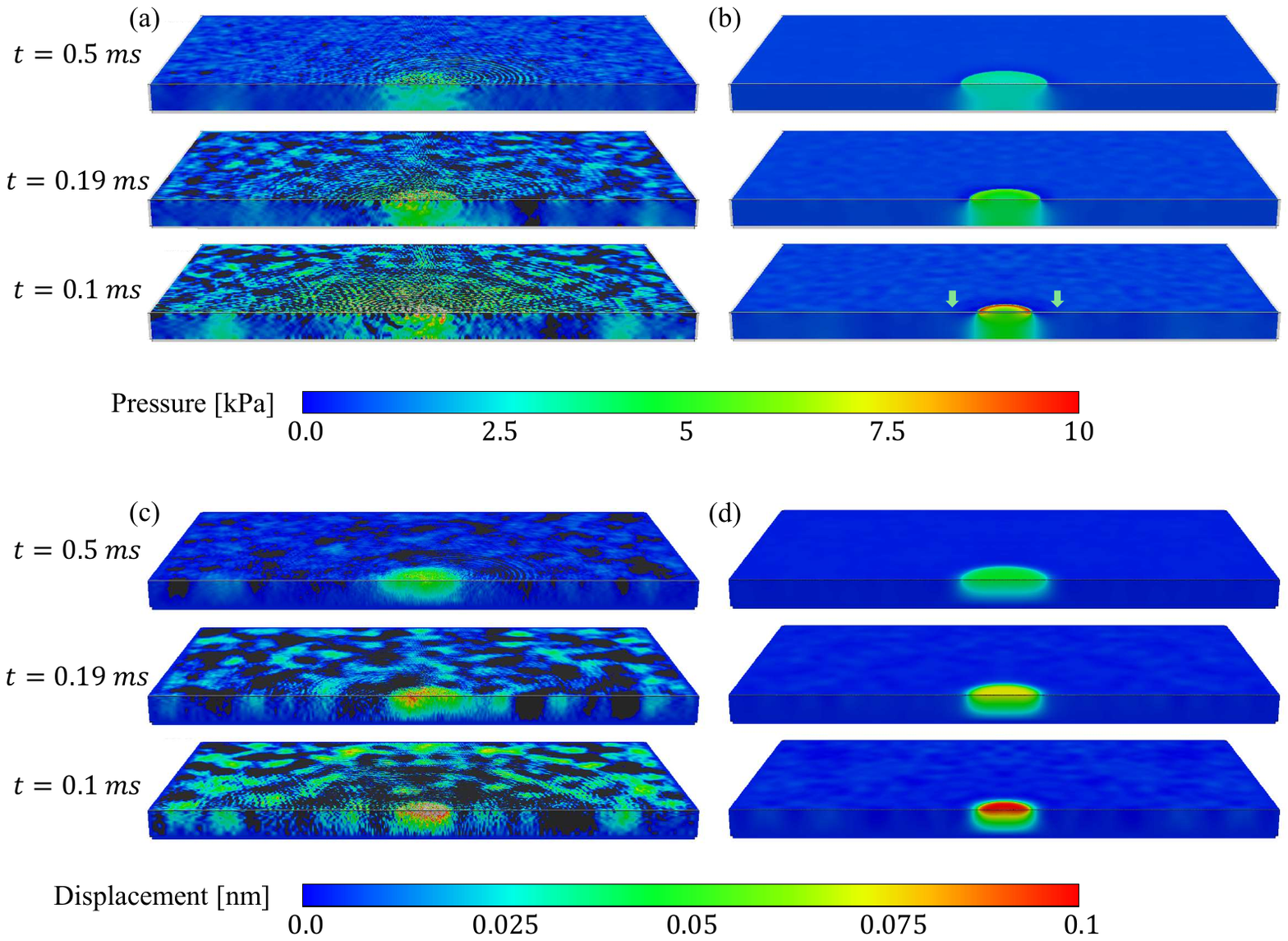}
    \caption{Spatial distributions of pressure and vertical displacement from SPH (a, c) and ANCM (b, d) simulations at (from bottom to top) $T=\qty{0.1}{\milli\second}$, \qty{0.19}{\milli\second}, and \qty{0.5}{\milli\second}.}
    \label{figure_results1}
\end{figure}

Figure \ref{figure_results1} compares the pressure and vertical displacement fields obtained from the SPH and ANCM simulations at $T=\qty{0.1}{\milli\second}$, $\qty{0.19}{\milli\second}$, and $\qty{0.5}{\milli\second}$, representing stages before, at, and after the peak force instant $T_p$, respectively. The SPH results exhibit pronounced numerical `noise' in the form of spurious oscillations, arising from particle dynamics. This issue is well documented in structural dynamic simulations of droplet impact \citep{Maheo1, Zhou1}, particularly at high strain rates, where the discretised liquid phase leads to discontinuous and noisy load transfer. By contrast, the ANCM results are smooth in both space and time, reflecting the continuous and noise-free loading prescribed by the analytical solution.

\begin{figure}
    \centering
    \includegraphics[width=0.7\linewidth]{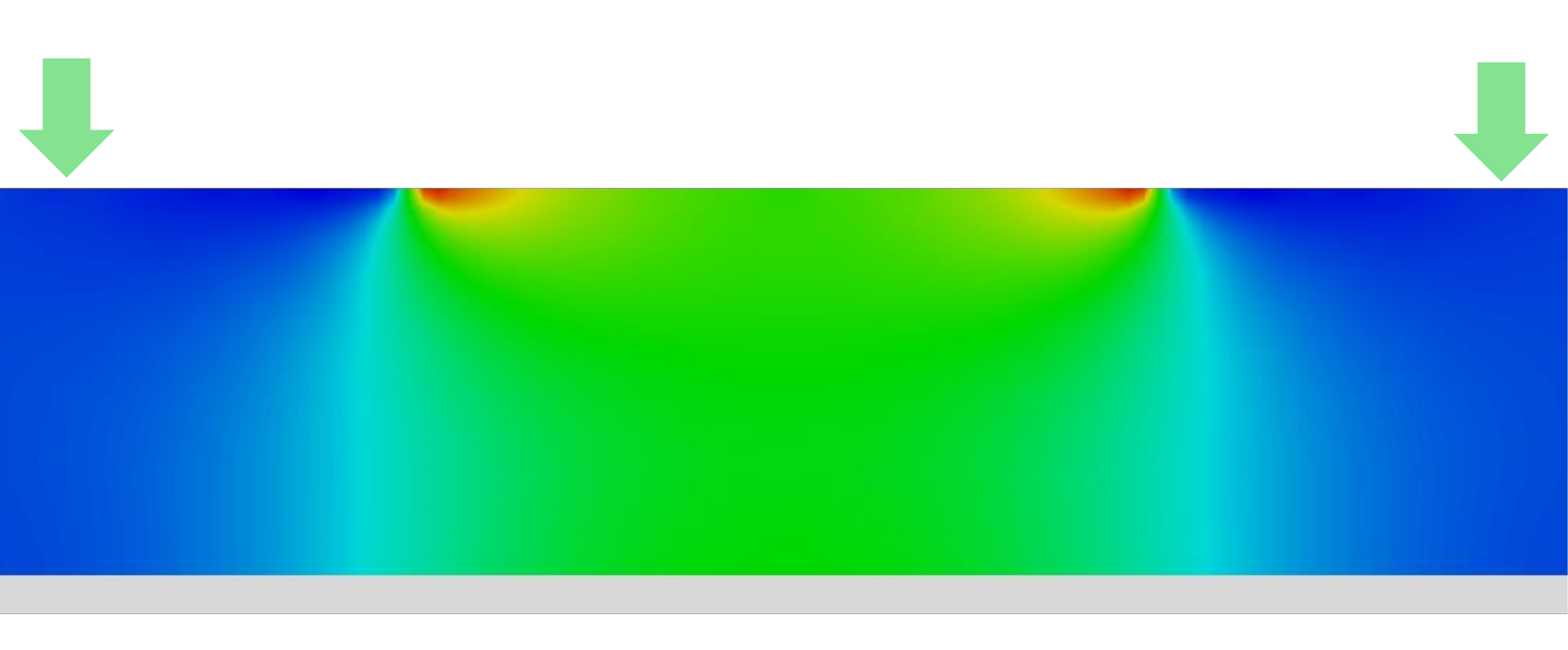}
    \caption{Zoom-in of the pressure distribution from the ANCM simulation at $T=\qty{0.1}{\milli\second}$ (same as the bottom plate in figure \ref{figure_results1}b) focusing on the central region of the aluminium plate over a length of \qty{4}{\milli\meter} (between the two arrows). The colormap and arrow positions correspond to those in figure \ref{figure_results1}b.}
    \label{figure_ring}
\end{figure}

\begin{figure}
    \centering
    \includegraphics[width=\linewidth]{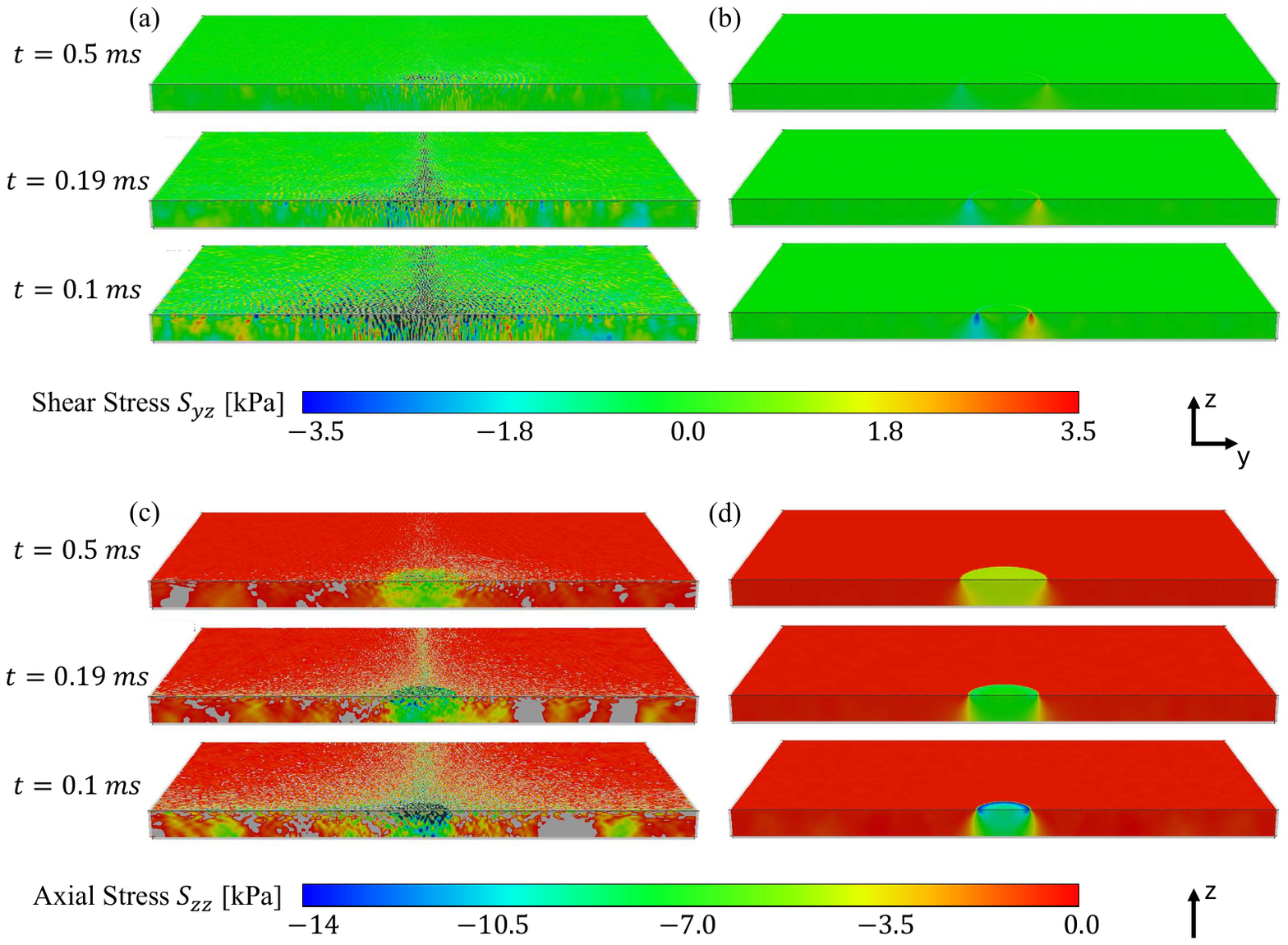}
    \caption{Spatial distributions of material shear stress $S23$ and vertical axial stress $S33$ from SPH (a, c) and ANCM (b, d) simulations at (from bottom to top) $T=\qty{0.1}{\milli\second}$, \qty{0.19}{\milli\second}, and \qty{0.5}{\milli\second}. For each stress component, the associated coordinates are indicated next to the colormap.}
    \label{figure_results2}
\end{figure}

Impact pressure distributions from both SPH (figure \ref{figure_results1}a) and ANCM (figure \ref{figure_results1}b) exhibit a clear ring-shaped pattern at early times for all three instants shown, consistent with experimentally observed ring-shaped erosion patterns \citep{Field1, Field3, Engel1, Brunton1}. Interestingly, this ring-pattern pressure does not penetrate through the plate thickness: beneath the surface, the pressure fields becomes dominantly centre-peaked over most of the depth (see figure \ref{figure_ring}). In contrast, the material (vertical) displacements fields from both SPH (figure \ref{figure_results1}c) and ANCM (figure \ref{figure_results1}d) consistently show a maximum displacement at the surface centre, decaying monotonically with radial distance. The peak displacement magnitude (excluding spurious oscillations) is of order $\textit{O}$(\SI{e-10}{\meter}), corresponding to approximately \SI{e-5}{\percent} of the droplet radius.

Figure \ref{figure_results2} shows the distributions of material stresses, including the shear stress $S23$ and the vertical axial stress $S33$, obtained from SPH and ANCM simulations at the same three instants ($T=\qty{0.1}{\milli\second}$, \qty{0.19}{\milli\second} and \qty{0.5}{\milli\second}. As before, the SPH results are affected by noticeable numerical noise. Nevertheless, when this is disregarded, good agreement is observed between the SPH and ANCM predictions. In particular, a ring-shaped pattern is evident in the vertical axial stress distributions (figures \ref{figure_results2}c and \ref{figure_results2}d), analogous to the surface pressure ring observed in figure \ref{figure_results1}b, and is captured more clearly by the ANCM results.

\subsection{Material Analysis: Quantitative Comparisons}\label{4.3}

\begin{figure}
    \centering
    \begin{subfigure}[b]{0.8\textwidth}
        \caption{}
        \includegraphics[width=\textwidth]{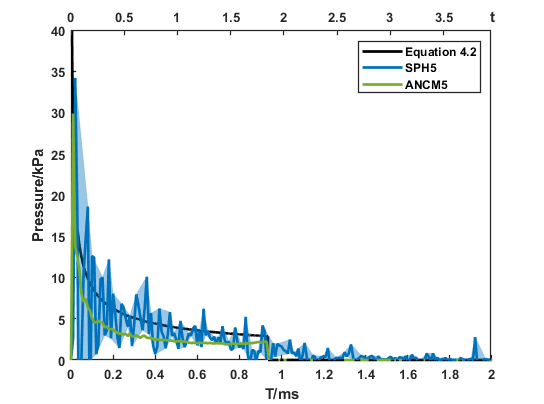}
        \label{figure_pre}
    \end{subfigure}
    \begin{subfigure}[b]{0.8\textwidth}
        \caption{}
        \includegraphics[width=\textwidth]{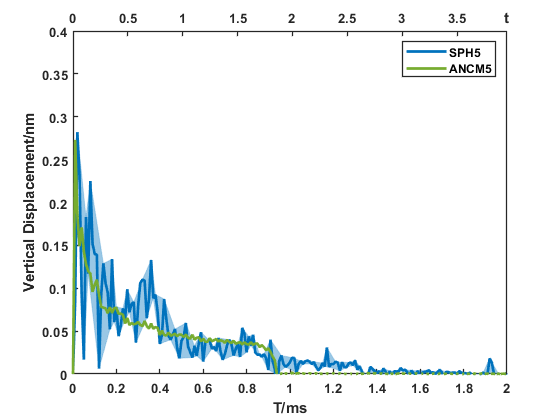}
        \label{figure_displacement}
    \end{subfigure}
    \caption{Temporal distributions of (a) pressure and (b) (vertical) displacement at the centre of the aluminium surface from SPH (shaded areas indicate numerical oscillation range) and ANCM simulations. In (a), the analytical solution from equation \ref{temp1} for the impact pressure on the plate surface is also superimposed.}
    \label{figure_results3}
\end{figure}

As a quantitative comparison, we examine the simulation results at the surface centre. Figure \ref{figure_results3} presents the temporal evolution of pressure and vertical displacement at the surface centre obtained from the SPH and ANCM simulations. For the pressure response (figure \ref{figure_pre}), the analytical pressure at the surface centre (\emph{i.e.} $r=0$) is also shown for reference,
\begin{equation}
  p(r=0,z=0,t) = \frac{3}{\pi \sqrt{3t}}.
\label{temp1}
\end{equation}
An interesting discrepancy is observed between equation \ref{temp1} and the ANCM results. This difference arises because equation \ref{temp1} represents the analytical loading for a rigid boundary, whereas the ANCM simulation accounts for pressure redistribution associated with deformation of the non-rigid aluminium plate during impact (see figure \ref{figure_results1}b). In effect, deformation of the solid acts as a cushion that disperses the applied pressure into the substrate. This mechanism may also help explain the transition from a surface ring-shaped pressure distribution to a centre-peaked pressure profile with depth, as observed in figure \ref{figure_results1}b. Comparing the SPH and ANCM pressure histories in figure \ref{figure_pre}, the SPH results exhibit pronounced numerical oscillations, which interestingly span the difference between the rigid-surface analytical prediction and the ANCM response. In contrast, the vertical displacement histories from SPH and ANCM show good agreement in overall trend, as illustrated in figure \ref{figure_displacement}.

\section{Discussion}\label{5}
This study extends the Wagner-type impact-flow framework in time to a full-duration analytic approximation by introducing two simplifying assumptions. One assumption addresses the wet-radius dynamics, $a(t)$, whose widely used analytical approximation, $\sqrt{3t}$, is over-simplified \citep{Egger1,Garcia2020} and requires moderation at late times \citep{Sanjay,Hao1}. The proposed modification of truncating the analytical framework at $r_{max}$ (from the more outboard $r_{sep}$), while not fully justified from first-principles physics, provides an effective means to achieve this moderation. As validated by numerical results, this represents the first analytical closure of characterising the non-monotonic evolution of the effective impact area, which progressively diminishes to zero as the initial normal impact transitions to lateral spreading.

Owing to the intrinsic limitations of the analytical framework, the flow is assumed to be inviscid and non-capillary. Consequently, the modified analytical solution is most appropriate for inertia-dominated regimes (\emph{i.e.} large Reynolds and Weber numbers), as discussed in section \ref{2.3}, and its applicability is limited for flows characterised by different non-dimensional parameters \citep{Sanjay}. For the application interest in the present study, leading-edge erosion on wind-turbine blades induced by water-droplet impact typically involves impact velocities in the range $50-100$ \unit{\meter\per\second} \citep{Nils2023}, corresponding to liquid Reynolds numbers $Re=\textit{O}(10^3)$ and Weber numbers $We=\textit{O}(10^4)$\footnote{These estimates assume droplet diameters of $1$ \unit{\milli\meter}.}. Under such conditions, viscous and capillary effects are negligible relative to inertial forces associated with the high-speed impact. We therefore conclude that the present analytical framework, together with the corresponding liquid-solid coupling model, is well suited for applications to impact erosion on wind-turbine blades.

The liquid-solid coupling model, termed the analytical-numerical coupled method (ANCM), is developed to analyse the response of solid materials to droplet impact by coupling an analytical solution for the liquid phase with a computer simulation for the solid response. Simulation results obtained using ANCM show good agreement with experimental measurements and with numerical predictions from the conventional smoothed particle hydrodynamics (SPH) method. This agreement holds for key quantities, including the impact force acting on the surface, as well as the pressure, stress and displacement fields within the solid, both temporally and spatially. Moreover, the proposed method effectively suppresses the numerical noise commonly encountered in dynamic FE simulations of contact problems \citep{Maheo1}, including conventional SPH-based droplet impact simulations. As a result, ANCM enables accurate analysis of material response upon impact, which is often compromised in SPH simulations by numerical noise. As an illustrative example, the ring-shaped pressure distribution on the contact surface following a single droplet impact\textemdash closely linked to erosion patterns on solid surfaces\textemdash has been extensively studied in fluid dynamics \citep{Riboux1, Philippi1, Negus1} and observed experimentally in materials research \citep{Field1, Field3, Engel1, Brunton1}. However, the spatial distribution and temporal evolution of this ring-pressure pattern beneath the contact surface have received comparatively little attention. ANCM simulations capture these subsurface pressure fields with high spatial and temporal resolution and smoothness, whereas SPH simulations struggle to achieve comparable clarity owing to numerical noise.

The most significant advantage of the proposed ANCM framework lies in its computational efficiency. By eliminating the need to explicitly simulate the liquid flow during droplet impact, the coupled approach substantially reduces the overall computational cost and, alternatively, allows significantly greater computational resources to be allocated to the solid-material analysis. According to the program monitor in ABAQUS, the computational time required by ANCM is approximately \qty{16}{\percent} of that required by the SPH method for the same mesh resolution. In addition, ANCM requires considerably fewer elements to achieve grid-independence results. When this factor is taken into account, the total computational time required to obtain a grid-independence solution using ANCM is reduced to only \qty{2.8}{\percent} of that required by the SPH method, while the associated memory usage is less than \qty{60}{\percent}.

To be precise about the quantified computational efficiency improvements reported above, it should be noted that a direct comparison between the analytical-numerical coupling approach and a fully numerical method is not strictly equivalent. The analytical solution provides only the impact loading on the contact surface (\emph{i.e.} $z=0$), without resolving the full droplet dynamics, such as droplet deformation \citep{Riboux2016,Josserand}, shock-induced oscillations \citep{Tsinghua1,Gonzalez-Avila_2024}, non-axisymmetric lamella rims \citep{ESA44,ESA45}, or droplet breakup and coalescence processes \citep{ESA42,ESA43}\textemdash all of which may give rise to secondary collisions and impacts at smaller scales\footnote{These secondary impacts also partly explain the numerical oscillations observed in SPH simulations.}. Such complex physical behaviours lie beyond the scope of the present analytical approximation. Accordingly, the reported computational efficiency should be interpreted in the context of solid material response analysis\textemdash such as impact erosion studies \citep{Mat1,Mat2}\textemdash where the primary quantity of interest is the transient loading imparted to the solid, rather than the detailed droplet morphology or fine-scale fluid dynamics. In this sense, the proposed ANCM framework genuinely achieves the reported reductions in computational cost by avoiding the simulation of fluid-phase dynamics that are not essential to the targeted engineering objectives.

\section{Conclusion}\label{6}
In this study, we derive an explicit, closed-form, full-duration analytic approximation for droplet impact onto a flat rigid solid surface in inertia-dominated regimes of large $Re$ and $We$. Building on this solution, we develop a novel fluid-solid coupling framework for investigating solid erosion induced by droplet impact by combining the analytical description of the liquid phase with numerical FE simulations of the solid response. In contrast to the conventional SPH approach, the proposed framework eliminates the need to explicitly simulate the droplet. Simulation results, including impact force and the distributions of pressure, stress and displacement, are validated against experimental data from the literature and against an SPH model developed in the present study. Key results are summarised as follows:
\begin{itemize}
    \item By modifying a theoretical framework of the liquid flow during droplet impact, we derive an approximate analytical closure for the impact and spreading phases in the explicit form. The modified analytical solution radially covers from the surface center to the pressure maximum radius $r_{max}$, and hence provides information for both peak radial pressure (and its position) and peak impact force on the surface.
    \item By combining the derived analytical solution of the liquid phase with a simulation of a solid material in FE analysis, we develop an analytical-numerical coupled method (ANCM) for studying the FSI problem of solid erosion induced by droplet impacts. The developed FSI model has two points of interest: it extends the theoretical framework of droplet fluid dynamics, which assumes rigid surface, into non-rigid solids of finite geometries; and for material analysis, it provides accurate and meshless fluid loading on the contact surface that cannot be otherwise achieved through numerical simulations.
    \item The proposed ANCM model demonstrates clear advantages over the conventional numerical SPH approach for simulating droplet impact problem. Numerical results obtained using SPH exhibit spurious oscillations, which, although mitigated by mesh refinement, nonetheless compromise the accuracy and clarity of the predictions. In contrast, the ANCM model produces results with negligible numerical oscillations and achieves grid independence at substantially lower mesh resolutions. Consequently, the ANCM model requires only \qty{2.8}{\percent} of the computational cost of the conventional SPH model. It therefore provides a more stable, accurate, and, most importantly, computationally efficient FSI framework for droplet impact erosion studies.
\end{itemize}

The developed ANCM model provides an alternative approach for studying liquid-solid impact problems in various engineering applications, such as leading-edge erosion on wind turbine blades, blade erosion in steam turbines, and rain erosion on aircraft. We expect that the analytical-numerical coupled approach allows reliable numerical predictions without the use of large computing resources or involved analyses.

\section*{Acknowledgements}
H. Hao acknowledges the Mechanical Engineering PhD Scholarship from the Department of Mechanical Engineering of Imperial College London.

\section*{Declaration of Interests}
The authors report no conflict of interest.

\bibliographystyle{jfm}
\bibliography{jfm-reference}

\end{document}